\documentclass[a4paper,usenatbib]{mnras}

\usepackage{newtxtext,newtxmath}

\usepackage[T1]{fontenc}
\usepackage{ae,aecompl}

\usepackage{graphicx}	
\usepackage{amsmath}	

\usepackage{float}
\usepackage{subfig}

\newcommand{\Msun}{{~\rm M_\odot}}
\newcommand{\Msunh}{{~\rm M_\odot/h}}

\newcommand{\mpc}{{~\rm Mpc}}

\newcommand{\mpch}{{~\rm Mpc/h}}
\newcommand{\eagle}{EAGLE}

\newcommand{\tng}{TNG100-1}

\newcommand{\lgalaxies}{\textsc{L-Galaxies}}

\newcommand{\eagleref}{Ref-L{\small 0100}N{\small 1504}}

\newcommand{\lgal}{LGal}
\newcommand{\lgalwo}{LGalw/oAGN}
\newcommand{\refeq}[1]{Eq.~\eqref{#1}}

\newcommand{\reftab}[1]{Table~\ref{#1}}
\newcommand{\reffig}[1]{Figure~\ref{#1}}
\newcommand{\red}[1]{\textbf{#1}}

\title[Reducing the scatter around SHMR]{Effectiveness of halo and galaxy properties in reducing the scatter in the stellar-to-halo mass relation}

\author[Pei et al.]{
Wenxiang Pei,$^{1,2,3}$
Qi Guo,$^{1,2,3}$\thanks{E-mail: guoqi@nao.cas.cn}
Shi Shao,$^{1}$
Yi He,$^{4}$
and Qing Gu$^{1}$
\vspace{0.1cm}
\\
$^{1}$Key Laboratory for Computational Astrophysics, National Astronomical Observatories, Chinese Academy of Sciences, Beijing 100101, China\\
$^{2}$Institute for Frontiers in Astronomy and Astrophysics, Beijing Normal University, Beijing 102206, China\\
$^{3}$School of Astronomy and Space Science, University of Chinese Academy of Sciences, Beijing 10049, China\\
$^{4}$Kapteyn Astronomical Institute, Postbus 800, NL-9700 AV Groningen, the Netherlands
}

\date{Accepted XXX. Received YYY; in original form ZZZ}

\pubyear{2015}

\begin{document}
\label{firstpage}
\pagerange{\pageref{firstpage}--\pageref{lastpage}}
\maketitle

\begin{abstract}
The stellar-to-halo mass relation (SHMR) is a fundamental relationship between galaxies and their host dark matter haloes. In this study, we examine the scatter in this relation for primary galaxies in the semi-analytic L-Galaxies model and two cosmological hydrodynamical simulations, \eagle{} and \tng{}. We find that in low-mass haloes, more massive galaxies tend to reside in haloes with higher concentration, earlier formation time, greater environmental density, earlier major mergers, and, to have older stellar populations, which is consistent with findings in various studies. Quantitative analysis reveals the varying significance of halo and galaxy properties in determining SHMR scatter across simulations and models. In \eagle{} and \tng{}, halo concentration and formation time primarily influence SHMR scatter for haloes with $M_{\rm h}<10^{12}\Msun$, but the influence diminishes at high mass. Baryonic processes play a more significant role in \lgal{}. For halos with $M_{\rm h} <10^{11}\Msun$ and $10^{12}\Msun<M_{\rm h}<10^{13}\Msun$, the main drivers of scatter are galaxy SFR and age. In the $10^{11.5}\Msun<M_{\rm h} <10^{12}\Msun$ range, halo concentration and formation time are the primary factors. And for halos with $M_{\rm h} > 10^{13}\Msun$, supermassive black hole mass becomes more important. Interestingly, it is found that AGN feedback may increase the amplitude of the scatter and decrease the dependence on halo properties at high masses. 

\end{abstract}




\section{Introduction}
\label{sec:intro}

In the standard $\Lambda$CDM cosmological framework, galaxies are firstly formed via the condensation of gas that has been cooled into the potential well of the host haloes \citep{1978MNRAS.183..341W}. Consequently, various properties of a galaxy such as the stellar mass, luminosity, or star formation rate are expected to exhibit significant correlations with the characteristics of the host dark matter halo.

One of the most fundamental galaxy-halo connections is the stellar-to-halo mass relation (hereafter referred to as SHMR) that connects the galaxy stellar mass to the virialized dark matter halo mass. This relation implies that the stellar mass of a galaxy is predominantly determined by its host halo mass. Numerous empirical observations and simulations substantiate this proposition  \citep[e.g.][]{2007ApJ...667..760Z,2010MNRAS.404.1111G,2010MNRAS.402.1796W,2012ApJ...752...41Y,2013MNRAS.428.3121M,2019MNRAS.488.3143B, 2020ApJ...899..137Z,2020A&A...634A.135G,2022MNRAS.512.6021N,2022A&A...664A..61S}. 

Nevertheless, the SHMR exhibits non-negligible scatter at a given halo mass, particularly for haloes with lower masses \citep[e.g.][]{2013MNRAS.428.3121M, 2013ApJ...770...57B, 2015ApJ...799..130R, 2017MNRAS.465.2381M, 2019MNRAS.482.3261K, 2021NatAs...5.1069C}. 
This means that using halo mass along could not predict the accurate stellar mass, and suggests the existence of secondary properties that might contribute to (at least part of) the scatter in the SHMR, such as halo formation time \citep[e.g.][]{2017MNRAS.465.2381M}.

The objective of this study is to explore the secondary properties contributing to the scatter of the SHMR. Many studies have focused on two closely related properties, halo concentration and formation time \citep{2002ApJ...568...52W, 2009ApJ...707..354Z, 2011MNRAS.415L..69J, 2014MNRAS.441..378L,2015MNRAS.450.1521C}. Haloes with high concentration have deep potential wells which can maintain more gas, thereby promoting star formation in their associated galaxies. Additionally, halos that formed earlier have had more time for accretion and star formation. For example, \citet{2017MNRAS.465.2381M} found that in the EAGLE simulation \citep{2015MNRAS.446..521S}, concentration and halo formation time substantially contribute to the scatter in the SHMR, particularly within the low halo mass regime. At fixed halo mass, haloes characterized by earlier formation times and higher concentrations tend to host more massive galaxies.
The impact of halo concentration and formation time on the scatter around the SHMR is also supported by other studies using either semi-analytic models (SAM) \citep{2013MNRAS.431..600W,2017MNRAS.470.3720T,2018ApJ...853...84Z,2023ApJ...959....5L} or hydrodynamical simulations \citep{2018MNRAS.480.3978A,2021NatAs...5.1069C}.
Some works find that other properties related to the formation time may also be able to modulate the SHMR, such as the environment \citep{2017MNRAS.470.3720T, 2018ApJ...853...84Z, 2021MNRAS.507.5320Z}, with denser environments exhibiting more massive galaxies at fixed halo mass, especially the low mass haloes; and the relaxation status \citep{2018ApJ...860....2G}, with a higher magnitude gap between the brightest central galaxy (BCG) and their brightest neighbours resulting in a massive BCG at fixed halo mass. On the other hand, \citet{2017MNRAS.465.2381M} found that halo properties unrelated to halo concentration and formation time (e.g., spin, sphericity, triaxiality, and substructure) demonstrate negligible impact on the scatter around the SHMR.

In addition to the halo properties, galaxy properties that are sensitive to certain formation processes could also affect the scatter of the SHMR. For example, the SHMR for star-forming (blue) and passive (red) galaxies could be different, though a consensus has yet to be reached \citep[see Section 6.1 of][for a review]{2018ARA&A..56..435W}. Several studies have observed that at a fixed halo mass, red galaxies tend to have larger stellar mass \citep{2016MNRAS.455..499L, 2016MNRAS.457.4360Z, 2018MNRAS.477.1822M,2019ApJ...874..114C,2022MNRAS.511.4900S}, while other investigations have suggested that blue galaxies are more massive \citep{2011MNRAS.410..210M,2013MNRAS.428.2407W,2015ApJ...799..130R,2016MNRAS.457.3200M,2019MNRAS.488.3143B,2020MNRAS.499.3578C,2021NatAs...5.1069C}.  \citet{2015MNRAS.447..298H} and \citet{2020MNRAS.499.2896T} have instead argued that no clear relationship exists between colour and the scatter in the SHMR. Some other dependence of the scatter around the SHMR is also proposed, such as galaxy age, \citep{2019MNRAS.482.3261K,2022MNRAS.511.4900S,2022ApJ...933...88O}, metallicity \citep{2022MNRAS.511.4900S,2022ApJ...933...88O}, cold gas \citep{2021NatAs...5.1069C}, and morphology \citep{2020MNRAS.499.2896T,2020MNRAS.499.3578C}. However, there is still no conclusive agreement on these dependencies. Challenges in reaching definitive conclusions are twofold: observationally, the direct measurement of halo mass is still challenging, and different observational endeavours typically rely on distinct halo mass estimation techniques, each accompanied by its own set of caveats and limitations (e.g., weak lensing \citep{2016MNRAS.457.4360Z}, satellite kinematics \citep{2011MNRAS.410..210M}, and group and cluster catalogues \citep{2020MNRAS.499.3578C,2022MNRAS.511.4900S}); theoretically, the treatment of sub-grid baryonic physics within numerical simulations is nontrivial, as the implementation, therefore, can significantly impact the interaction between galaxies and haloes.

Although many efforts have been made to study the link between stellar mass and various halo and galaxy properties, the quantitative evaluation and comparison of their impact on reducing scatter around the SHMR is lacking. In most previous studies such an analysis was conducted focusing solely on halo properties \citep{2017MNRAS.465.2381M, 2020MNRAS.491.5747M,2020MNRAS.493..337B}. Here, we perform a comprehensive analysis not only on the general trend between the stellar mass and different halo and galaxy properties but also to evaluate and compare how effectively these properties can minimize scatter around the SHMR. Additionally, we will examine the discrepancies among the advanced semi-analytic model \lgalaxies{} \citep{2015MNRAS.451.2663H} and two cosmological hydrodynamical simulations, the Evolution and Assembly of Galaxies and their Environments \citep[\eagle,][]{2015MNRAS.446..521S,2016A&C....15...72M} and \emph{The Next Generation Illustris Simulations} \citep[IllustrisTNG,][]{2018MNRAS.475..648P,2018MNRAS.475..624N,2018MNRAS.477.1206N,2018MNRAS.475..676S,2018MNRAS.480.5113M}. The analysis focuses on the influence of different halo and galaxy characteristics such as halo concentration, formation time, major halo mergers, star formation rate, black hole mass, cold gas mass, and others. Specifically, we examine the impact of AGN feedback, crucial for regulating galaxy formation in higher masses, by implementing \lgalaxies{} on N-body cosmological simulations without AGN feedback to evaluate its effect on the SHMR. We employ traditional statistical techniques and machine learning to systematically investigate the correlation coefficients between the scatter in SHMR and halo/galaxy properties.

This paper is organized as follows. Section~\ref{sec:data} reviews the simulations used in this work and describes our sample selection; Section~\ref{sec:results} presents our results on the dependence of various halo/galaxy properties on the scatter in SHMR; Section~\ref{sec:AGN} discusses the influence of AGN feedback; we conclude with a summary and discussion in Section~\ref{sec:conclusion}.

\section{Simulation data and methods}
\label{sec:data}
We employ a semi-analytic   galaxy catalogue, and two cosmological hydrodynamical simulations, \eagle{} \citep{2015MNRAS.446..521S} and \tng{} \citep{2017MNRAS.465.3291W}, to explore the origin of the scatter in SHMR. The semi-analytic galaxy catalogue is generated by implementing the galaxy formation model \lgalaxies{} \citep{2015MNRAS.451.2663H}  onto merger trees extracted from two large dark matter and gravity only $N$-body cosmological simulations, the Millennium Simulation \citep[MR;][]{2005Natur.435..629S} and the Millennium-II Simulation \citep[MRII;][]{2009MNRAS.398.1150B}. A summary of the simulations is provided in \reftab{tab:simulations}. In both semi-analytical and hydrodynamical models, the gas condenses in the centre of the hierarchically assembled dark halos through shock heating and cooling. The mass exchange between stellar components and condensed gas is regulated by star formation and various feedback. In general, galaxies can grow through in situ star formation and mergers. Star formation has been shown to be the primary growth mechanism in most galaxies except for the very massive ones, where star formation is suppressed by AGN feedback. For the most massive galaxies, mergers and merger-induced starbursts may be the sole channel for the growth of the galaxy. 

The properties of the explored galaxy are closely linked to SFR, cold gas, and AGN feedback. We provide a summary of the simulations and their treatment of SFR and AGN feedback.

\begin{table}
	\centering
	\caption{Simulations and sample selection. The columns from left to right represent: the name of the simulation, the number of central galaxies, halo mass cut, dark matter particle mass, and baryonic particle mass.}
	\label{tab:simulations}
	\begin{tabular}{lcccr} 
		\hline
              & $N_{\rm cen}$ & $M_{\rm h,cut}[\Msun]$ & $M_{\rm dm}[\Msun]$ & $M_{\rm bar}[\Msun]$ \\
		\hline
		MR & 531,953 & $10^{12}$ & $1.4\times 10^{9}$ & -\\
		MRII & 74,489 & $10^{10.6}$ & $1.14\times 10^{7}$ & -\\
		\eagle{} & 21,387 & $10^{10.6}$ & $1.81\times 10^{8}$ & $9.7\times 10^{6}$\\
            \tng{} & 31,990 & $10^{10.6}$ & $1.4\times 10^{8}$ & $7.5\times 10^{6}$\\
		\hline
	\end{tabular}

\end{table}

\subsection{L-Galaxies semi-analytic model}
\label{sec:lgalaxy} 
The semi-analytic model of galaxy formation, \lgalaxies, was initially introduced in \citet{2001MNRAS.328..726S} and subsequently refined in a series of versions \citep{2004MNRAS.349.1101D,2006MNRAS.365...11C,2011MNRAS.413..101G}. These models share similar philosophies which include physical prescriptions for baryonic processes such as shock heating, gas cooling, star formation, supernova feedback, formation and growth of supermassive black holes, AGN feedback, metal enrichment, etc. In this study, we adopt the model developed in \citet{2015MNRAS.451.2663H}, which introduced modifications to the treatment of the reincorporating wind ejecta, star formation thresholds, environmental stripping, and the radio mode feedback. The model is fitting on the observed galaxy stellar mass function and passive fractions of galaxies at redshift $0 \le z \le 3$. They execute the model on the merger trees extracted from the Millennium Simulation \citep[MR;][]{2005Natur.435..629S} and the Millennium-II Simulation \citep[MRII;][]{2009MNRAS.398.1150B}, and rescale the simulations to those with the first-year \emph{Planck} cosmology parameters\citep{2014A&A...571A..16P}: $\sigma_8=0.829$, $H_0=67.3 \rm ~km ~s^{-1}~Mpc^{-1}$, $\Omega_\Lambda=0.685$, $\Omega_{\rm m}=0.315$, $\Omega_{\rm b}=0.0487\ (f_b=0.155)$ and $n=0.96$. Both MR and MRII trace $2160^3$ dark matter particles from redshift $\sim$56.4 to 0. The MR and MRII were carried out in periodical boxes of 480.279\mpch{} and 96.0558\mpch{} on each side after rescaling, respectively. The corresponding dark matter particle masses are $9.61\times10^{8}\Msunh$ and $7.69\times10^{6}\Msunh$. Details about the simulations and semi-analytical models can be found in  \citet{2015MNRAS.451.2663H} and reference therein.

In \lgalaxies{}, There are two modes to form stars: the quiescence mode and the burst mode. In the quiescence mode where no merger occurs, the star formation rate is proportional to the amount of cold gas mass when its surface density surpasses a specific threshold,
\begin{equation}
    \dot{M}_{*}=\alpha_{\rm{SF}} \frac{\left(M_{\rm{gas}}-M_{\rm{crit}}\right)}{t_{\rm{dyn}, \rm{disk}}},
\end{equation}
where $\alpha_{\rm{SF}} = 0.030$ is a free parameter, $M_{\rm gas}$ is the total mass of cold gas, and $t_{\rm{dyn}, \rm{disk}}$ is the dynamical time of the disk. $M_{\rm{crit}}$ is a threshold mass converted from the critical surface density to form stars \citep{1996MNRAS.281..487K}:
\begin{equation}
    M_{\text {crit }}=M_{\text {crit }, 0}\left(\frac{V_{200 \rm{c}}}{200 \rm{km} \rm{s}^{-1}}\right)\left(\frac{R_{\text {gas }}}{10 \rm{kpc}}\right),
\end{equation}
where $M_{\text {crit }, 0} = 0.24 \times 10^{10} \Msun\, {\rm pc ^{-2}}$ is another free parameter, $V_{200 \rm{c}}$ is the virial velocity of the halo, and $R_{\text {gas }}$ is the cold gas disk radius. In the burst mode where merger occurs, the stellar mass formed is determined by the “collisional starburst” formulation \citep{2001MNRAS.320..504S}: 
\begin{equation}
    M_{*, \text { burst }}=\alpha_{\mathrm{SF}, \mathrm{burst}}\left(\frac{M_1}{M_2}\right)^{\beta_{\mathrm{SF}, \mathrm{burst}}} M_{\text {cold}},
\end{equation}
where $M_1\,<\,M_2$ is the baryonic mass of the two galaxies, and $M_{\rm cold}$ is total cold gas mass. $\alpha_{\mathrm{SF}, \mathrm{burst}} = 0.72$ and $\beta_{\mathrm{SF}, \mathrm{burst}}=2.0$ are two free parameters. 

AGN feedback plays a crucial role in suppressing star formation in massive galaxies, influencing the shape of the SHMR in the high-mass regime. 
In the \lgalaxies{} model \citep{2015MNRAS.451.2663H}, a black hole with mass 0 is seeded when a halo is formed. The black hole can then grow through two modes: \textit{Quasar mode} and \textit{Radio mode}. The former, a primary SMBH growth mechanism, occurs during galaxy mergers, where violent merger processes funnel a substantial amount of cold gas into the black hole \citep{2021RAA....21..212Z}. In contrast, the \textit{Radio mode} refers to Bondi accretion \citep{1944MNRAS.104..273B} 
 of hot gas from the host halo. The mass accretion rate is determined both by the SMBH mass and the hot gas mass:
\begin{equation}
    \dot{M}_{\rm acc}=k_{\rm AGN}\frac{M_{\rm hot}}{10^{11}\Msun}\frac{M_{\rm BH}}{10^{8}\Msun},
\end{equation}
where $k_{\rm AGN} = 5.3\times 10^{-3}$, $M_{\rm hot}$ and $M_{\rm BH}$ are the hot gas mass and the central SMBH mass, respectively. Although \textit{Radio mode} accretion contributes minimally to the final BH mass, it is assumed to be strongly correlated to the AGN feedback. The thermal energy from AGN feedback is given by:
\begin{equation}
    \dot{E}_{\rm thermal}=\eta\dot{M}_{\rm acc}c^2,
\end{equation}
where $\eta=0.1$ is the efficiency parameter in the fiducial model and $c$ is the speed of light. The feedback prevents gas from cooling onto the galaxies and consequently suppresses further star formation.

To examine the impact of AGN feedback, we rerun the model presented in \cite{2015MNRAS.451.2663H} on both MR and MRII simulations with AGN feedback switched off, which is referred to as \lgalwo{}. The simulation with default \lgalaxies{} model is referred to as \lgal{}.
Note that due to the limited box size of MRII and the lower resolution of MR, we choose halos of mass [$10^{10.6}$, $10^{12}$]$\Msun$ from MRII and [$10^{12}$,$\infty$]$\Msun$ from MR for the analysis. LGal and LGalw/oAGN refer to galaxy catalogues on these combined halo merger trees, unless stated otherwise.
The galaxy formation model parameters are adjusted to reproduce the stellar mass function from z = 0 to 3, both at high masses based on MR and at low masses based on MRII. However, it should be noted that the galaxy formation model and parameters are not calibrated according to the scatter around the SHMR. (see Section~\ref{sec:smhm} for details).

\subsection{EAGLE simulation}
\label{sec:eagle}

\eagle{} project constitutes a suite of hydrodynamical simulations designed to track the formation of galaxies and supermassive black holes, following the evolution of gas, stars and dark matter in cosmologically representative volumes within a standard $\Lambda$CDM universe \citep{2015MNRAS.446..521S,2016A&C....15...72M}. These simulations utilized a modified version of the N-Body Tree-PM smoothed particle hydrodynamics (SPH) code gadget 3 \citep{2005MNRAS.364.1105S}. The sub-grid physics includes element-by-element radiative cooling for 11 elements, star formation, stellar mass loss, energy feedback from star formation, gas accretion and mergers of supermassive black holes, and AGN feedback. These subgrid models were calibrated to reproduce the stellar mass function, galaxy sizes, and the relation between galaxy stellar masses and supermassive black hole masses at the present day. A more detailed description of the implemented baryonic processes in \eagle{} is provided in \citet{2015MNRAS.446..521S}.
 
For our analysis,  we adopt the \eagle{} \eagleref{} (hereafter \eagle{}) simulation, which traces $2\times1504^3$ dark matter and baryonic particles within a periodic box of 100\mpc{} on each side. The gas particle and dark matter particle masses are $1.81\times 10^6\Msun$ and $9.70\times 10^6\Msun$, respectively. The cosmological parameters adopted in the simulation are from Planck Collaboration XVI \citep{2014A&A...571A..16P}: $\sigma_8=0.8288$, $H_0=67.77 \rm ~km ~s^{-1}~Mpc^{-1}$, $\Omega_\Lambda=0.693$, $\Omega_{\rm m}=0.307$, $\Omega_{\rm b}=0.04825$ and $n=0.9611$.

In the \eagle{} simulation, the star formation rate is determined by pressure rather than density, following the equation:
\begin{equation}
\label{sfr_eagle}
   \dot{M}_*=m_{\mathrm{g}} A\left(1 \mathrm{M}_{\odot} \mathrm{pc}^{-2}\right)^{-n}\left(\frac{\gamma}{G} f_{\mathrm{g}} P\right)^{(n-1) / 2},
\end{equation}
where $m_{\rm g}$ is the mass of a gas particle, $\gamma = 5/3$ is the ratio of specific heats, $G$ is the gravitational constant, $f_g$ is the mass fraction in gas, and $P$ is the total pressure. $A = 1.515\times 10^{-4}\Msun \rm yr^{-1} kpc^{-2}$ and $n = 1.4$ are two free parameters, while for $n_{\rm H} > 10^{3} \rm cm^{-3}$, $n = 2$. The probability of a gas particle converted to a star particle during a time step $\Delta t$ is given by min($\dot{m_{\ast}}\Delta t/m_{\rm g}, 1$). Additionally, a metallicity-dependent density threshold and a temperature threshold for gas particles are imposed to regulate star formation.

In the \eagle{} model, a black hole with seed mass $M_{\rm BH, seed} = 10^{5}\Msunh$ is placed in the centre of the host halo when its mass exceeds $10^{10}\Msunh$ for the first time. BH growth can occur through BH-BH mergers and gas accretion. The gas accretion rate is given by:
\begin{equation}
    \dot{M}_{\rm acc}={\rm min}(\dot{M}_{\rm Edd}, \dot{M}_{\rm bondi}\times {\rm min}(C^{-1}_{\rm visc}(c_{\rm s}/V_{\rm \phi})^3, 1)),
\end{equation}
where $\dot{M}_{\rm bondi}$ is the Bondi accretion rate and $\dot{M}_{\rm Edd}$ is the Eddington rate. The factor $(c_{\rm s}/V_{\rm \phi})^3/C_{\rm visc}$ is the ratio between Bondi and the viscous time-scales \citep{2015MNRAS.454.1038R}. Here $V_{\rm \phi}$ is the rotation speed of the gas around the BH and $C_{\rm visc}$ is a free parameter. The thermal energy injection rate is assumed to depend on the gas accretion rate:
\begin{equation}
    \dot{E}_{\rm thermal}=\epsilon_{\rm f}\epsilon_{\rm r}\dot{M}_{\rm acc}c^2,
\end{equation}
where $\epsilon_{\rm f}$=0.15 and $\epsilon_{\rm r}$ = 0.1. The details of the model are referred to \citet{2015MNRAS.446..521S}.

\subsection{IllustrisTNG simulation}
\label{sec:TNG}

\textit{IllustrisTNG} comprises a suite of cosmological hydrodynamical simulations focused on galaxy formation. They were carried out using the moving-mesh code AREPO \citep{2010MNRAS.401..791S} with updated treatments of the formation, growth and feedback of black holes, galactic winds, stellar evolution and chemical yields and enrichment. In addition, they included magnetic fields based on the previous ideal
magneto-hydrodynamics \citep{2011MNRAS.418.1392P, 2013MNRAS.432..176P, 2014ApJ...783L..20P}. A comprehensive description of the full physics involved can be found in \citet{2017MNRAS.465.3291W} and \citet{2018MNRAS.473.4077P}. 
In this study, we utilize the TNG100-1 simulation for comparison with \lgalaxies{} and \eagle{} simulations. TNG100-1 traces $2 \times 1820^3$ dark matter and baryon particles within a periodic box of $L = 75\mpch$ on each side. The mass resolution is $9.4\times10^5 \Msunh$ and $5.1\times10^6 \Msunh$ for dark matter and baryonic particles, respectively. They adopted cosmological parameters from more recent Planck Collaboration \citep{2016A&A...594A..13P}, which is different to those adopted by \lgalaxies{} and \eagle{}: $\sigma_8=0.8159$, $H_0=67.74 \rm ~km ~s^{-1}~Mpc^{-1}$, $\Omega_\Lambda=0.6911$, $\Omega_{\rm m}=0.3089$, $\Omega_{\rm b}=0.0486$ and $n=0.9677$.

In the \tng{} simulation, stars form stochastically in gas particles when their densities exceed a star formation threshold of $n_{\rm H} \approx 0.1 \rm cm^{-3}$. The star formation rate is given by :
\begin{equation}
    \dot{M}_*=\frac{\rho_{\mathrm{c}}}{t_{\star}}-\beta \frac{\rho_{\mathrm{c}}}{t_{\star}}=(1-\beta) \frac{\rho_{\mathrm{c}}}{t_{\star}},
\end{equation}
where $\rho_{\mathrm{c}}$ is the density of cold clouds, $\beta = 0.1$ is the mass fraction of short-lived massive stars, and $t_{\star} = 2.2\, \rm Gyr$ is the time-scale for star formation. 

In \tng{} simulation, an initial mass of $M_{\rm BH, seed} = 8\times10^{5}\Msunh$ is placed into a host halo when its mass reaches $M_{\rm host}=5\times10^{10}\Msunh$ for the first time. The SMBH accretion rate in the model is the minimum value between the Eddington accretion rate and the Bondi-Hoyle-Lyttleton accretion rate:
\begin{equation}
    \dot{M}_{\rm acc}=min(\dot{M}_{\rm Edd}, \dot{M}_{\rm bondi}).
\end{equation}
The model incorporates two AGN feedback modes which depend on the accretion rate. At low-accretion rates ($\chi=\dot{M}_{\rm bondi}/\dot{M}_{\rm Edd} \leq 0.1$), the AGN-driven wind feedback takes effect by blowing out the surrounding gas kinematically:
 \begin{equation}
    \dot{E}_{\rm kin}=\epsilon_{\rm f,kin}\dot{M}_{\rm acc}c^2,
\end{equation}
where a maximum value of $\epsilon_{\rm f,kin} = 0.2$ is set. This involves adding momentum in a random direction by kicking each gas particle cell in the feedback region. At high-accretion rates ($\chi \ge 0.1$), thermal feedback is activated with an injection rate:
 \begin{equation}
    \dot{E}_{\rm thermal}=\epsilon_{\rm f}\epsilon_{\rm r}\dot{M}_{\rm acc}c^2,
\end{equation}
where $\epsilon_{\rm f}$=0.1 and $\epsilon_{\rm r}$ = 0.2, slightly larger than the parameters used in \eagle{}. For more details of the model, readers are referred to \citet{2017MNRAS.465.3291W}.

Previous works reveal that although both hydrodynamical simulation and \lgalaxies{} can reproduce the overall galaxy stellar mass function, they might differ in detail. For example, \citet{2016MNRAS.461.3457G} investigate the galaxies in \eagle{} and \lgalaxies{} and find that \lgalaxies{} predict more star-forming galaxies and higher star formation rate density than \eagle{}. By running \lgalaxies{} on the DMO merger trees of \tng{}, \citet{2021MNRAS.502.1051A} find significant differences in the sSFR and cold gas component between semi-analytical model and hydrodynamical simulation, which are primarily due to the different stellar feedback and AGN feedback. 

In all these three simulations, the halo virial mass is defined as the total mass enclosed within a sphere whose mean density is 200 times the critical density of the Universe, denoted as $M_{\rm 200c}$. It is worth noting that in \lgalaxies{}, the dark matter properties are directly inherited from dark matter only (DMO) merger trees, which are unaffected by baryonic processes. However, in \eagle{} and \tng{}, the dark matter properties are derived from hydrodynamical simulations, rather than the DMO versions used in \citet{2017MNRAS.465.2381M}. Therefore, these halo properties could be influenced by baryonic processes.
Galaxy properties such as total stellar mass, star formation rate (SFR), and black hole mass are determined by corresponding quantities within subhalos, accessible directly from the catalogue. The cold gas mass can also be obtained from the \lgal{} and \lgalwo{} output catalogue. In \eagle{}, we use the fitting formula from Appendix A in \citet{2013MNRAS.430.2427R} to define cold gas mass, linking cosmological simulations with full radiative transfer calculations to predict the neutral gas fraction. In \tng{}, cold gas mass represents the total mass of star-forming particles plus other gas particles adjusted by the "NeutralHydrogenAbundance" factor.

The merger trees in these simulations are all constructed following the method outlined in \citet{2005Natur.435..629S}. In practice, at each snapshot, the FOF and SUBFIND algorithms \citep{2001MNRAS.328..726S} are used to identify groups and subhalos. Each subhalo is associated with one and only one descendant. Then the merger trees are constructed by linking each subhalo with its descendant at the following snapshot. For further details regarding the construction of merger trees, we refer readers to \citet{2005Natur.435..629S} and the references therein.

We focus on central galaxies in this study. Selecting for host halo masses above $M_{\rm h}>10^{10.6}\Msun$, we find a total of 21,387 central galaxies in \eagle{} and 31,990 in \tng at z=0. Central galaxies are chosen from \lgal{} (\lgalwo{}) on MRII and MR based on their host halo masses of [$10^{10.6}$, $10^{12}$]$\Msun$ and [$10^{12}$,$\infty$]$\Msun$, resulting in 74,489 and  531,953 central galaxies at z = 0, respectively.

\subsection{Stellar mass vs. halo mass in different simulations}
\label{sec:smhm}
\begin{figure}
\includegraphics[width=\columnwidth]{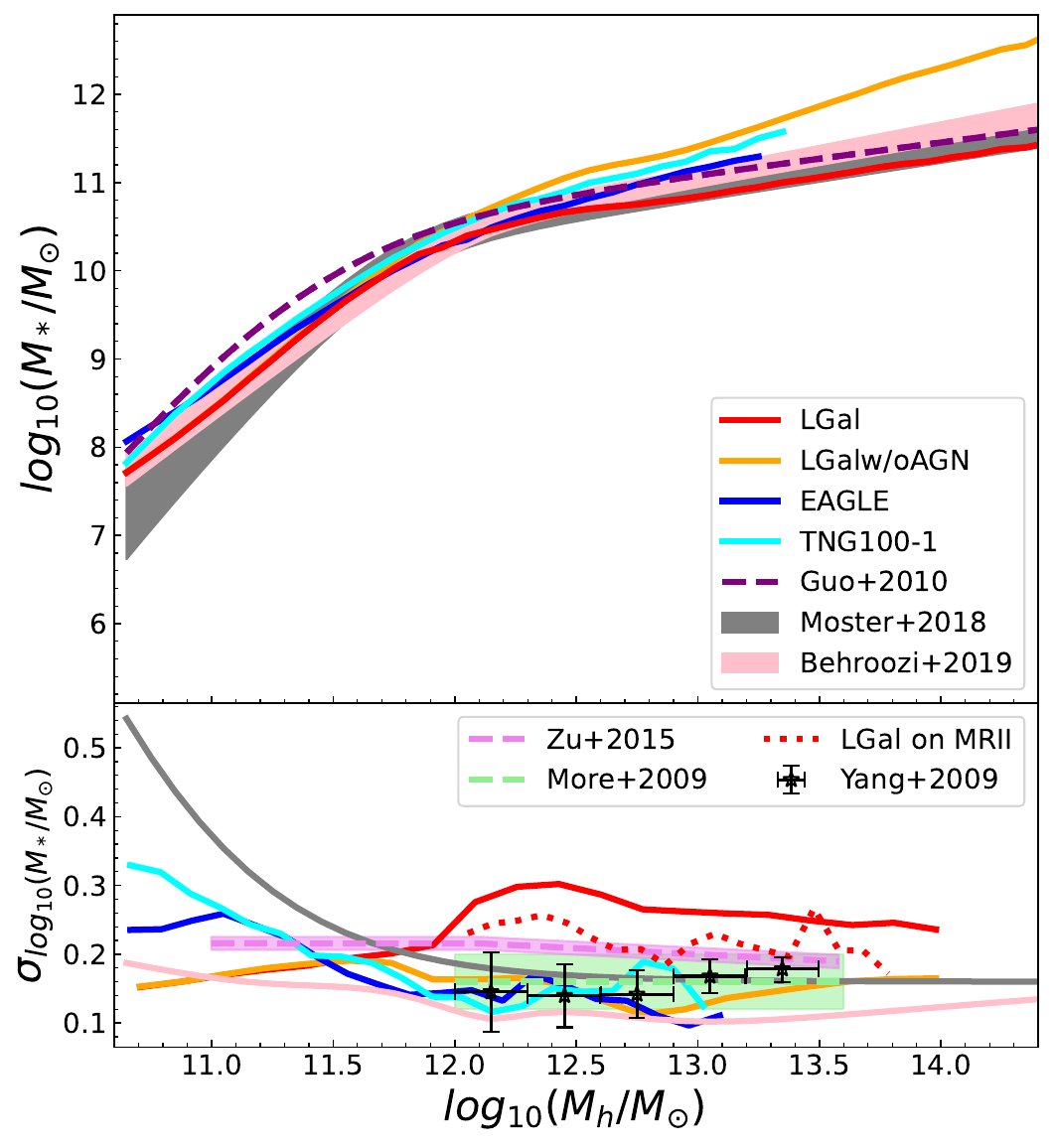}
\caption{\textit{Top panel:} the median values of the stellar-halo mass relation (SHMR) in different simulations. Red (Orange) lines show the results of LGalaxies with (without) the inclusion of the AGN model. Blue and cyan lines show the results from hydrodynamical simulations, \eagle{} and \tng{}, respectively. For comparison, the dashed line shows the result from the semi-analytical model by \citet{2010MNRAS.404.1111G} and the results (shaded regions) obtained from abundance matching techniques \citep{2018MNRAS.477.1822M,2019MNRAS.488.3143B}. We only show the data points with their mass bins containing at least 20 galaxies. \textit{Bottom panel:} the 1$\sigma$ scatter of stellar mass as a function of halo mass. The red dotted line shows the result from LGal\_on\_MRII. Observational results \citep{2009ApJ...695..900Y,2009MNRAS.392..801M,2015MNRAS.454.1161Z} are presented for comparison.
}
\label{fig:smhm}
\end{figure}

The top panel of \reffig{fig:smhm} depicts the median stellar mass as a function of halo mass for central galaxies in \lgalaxies{}, \eagle{}, and \tng{} simulations. The central galaxies in \lgalaxies{} consist of those in low mass haloes ($M_{\rm h}<10^{12}\Msun$) in MRII (\lgalaxies{} on MRII) and those in higher mass haloes ($M_{\rm h}>10^{12}\Msun$) in MR (\lgalaxies{} on MR). Notably, the SHMR in \lgal{} with AGN feedback overlaps with the result in \lgalwo{} without AGN feedback for low-mass haloes ($M_{\rm h}<10^{12}\Msun$). However, the difference in SHMR between these two models increases with increasing halo mass, with the stellar mass of galaxy clusters ($M_{\rm h}\sim10^{14}\Msun$) in \lgalwo{} being 1 dex more massive than their counterparts in \lgal{}. This result is consistent with previous works such as \citet{2022hxga.book....5H}, suggesting the efficiency of AGN feedback in suppressing star formation is the strongest in galaxy groups or galaxy clusters. 

Comparing the semi-analytic catalogue and hydrodynamical results, it is evident that the SHMR in hydrodynamical simulations (\eagle{} and \tng) are consistent with \lgal{} predictions at $M_{\rm h}\sim10^{12}\Msun$, but systematically higher than \lgal{} both at higher and lower mass ranges. At low masses ($M_{\rm h}<10^{11}\Msun$), the excess could be attributed to some backsplash galaxies that once resided in more massive haloes thus their halo being disrupted. The discrepancy is more extreme in the high-mass regime. For example, for a given halo with $M_{\rm h}\sim10^{13}\Msun$, the stellar mass in \tng{} or \eagle{} is roughly 3 times larger than \lgal{} predictions. This discrepancy suggests that AGN feedback models employed in hydro simulations are less effective for cluster-mass haloes than \lgal{}. 
Despite such differences, SHMR in all simulations aligns with those obtained using the abundance matching method\citep[][]{2018MNRAS.477.1822M,2019MNRAS.488.3143B}. 

The bottom panel of \reffig{fig:smhm} illustrates the 1$\sigma$ scatter of the SHMR for \lgal{}, \lgalwo{}, \tng{}, \eagle{} and abundance matching methods. Generally, the scatter varies with the host halo mass at low masses and becomes more constant at $M_{\rm h}>10^{12}\Msun$. For comparison, many abundance matching methods assume a constant scatter \citep[e.g. ][]{2010MNRAS.404.1111G,2012ApJ...744..159L,2013MNRAS.428.3121M,2016MNRAS.459.3251V}. Detailed comparison shows that in \tng{} and \eagle{}, the scatter decreases with increasing host halo masses, while in \lgal{} it increases with halo mass, reaching the maximum value $M_{\rm h}\sim 10^{12.2}\Msun$, and then decreases. The decreasing scatter feature with halo mass in hydrodynamical simulations is not influenced by the limited size of simulation boxes (refer to details in the Appendix~\ref{app:conv}). The scatter is larger in \lgal{} at high masses and smaller at low masses compared to \tng{} and \eagle{}.  The large scatter at low masses could be attributed to baryonic effects on the dark halo in \tng{} and \eagle{}, which are absent in semi-analytical approaches. At high masses, small scatter in \tng{} and \eagle{} could result from less effective AGN feedback, as observed in \lgalwo{}. We note that the scatter in \lgalaxies{} on MRII as denoted by the red dotted curve is much smaller than that of \lgalaxies{} on MR at high masses and is more in line with the scatters found in \tng{} and \eagle{}. This could be attributed to different SMBH growth histories which are closely related to mergers and thus sensitive to the resolution of the simulation. Such reliance on the resolution of the scatter is not present in \tng{} (Appendix~\ref{app:conv}). In the mass ranges between $10^{11}\Msun$ and $10^{12}\Msun$, where AGN feedback is not effective and the MR resolution is marginally sufficient, the scatter is indeed very similar between the \lgalaxies{} on MR and \lgalaxies{} on MRII (see Appendix~\ref{app:shmr}.). Since \lgalaxies{} are calibrated using MR for massive objects, high mass galaxies from MRII are not included in this study. In the absence of AGN feedback, the SHMR and its scatter are similar between \lgalwo{} on MR and MRII across the entire mass ranges.

The model scatter is in good agreement with various observational results for most simulations. The only exception is the results based on MR, as shown by the red curve at high masses, which is much larger than the observed ones. However, the results of \lgalaxies{} on MRII are in good agreement with \citet{2015MNRAS.454.1161Z}. This demonstrated that the diversity of SMBH growth in \lgalaxies{} on MR might have been overpredicted due to the poor resolution. The scatter in the SDSS group catalogue \citep{2009ApJ...695..900Y} and in groups whose mass is estimated with satellite kinematics \citep{2009MNRAS.392..801M} has a mean value of 0.15 dex, similar to those used in the abundance matching method \citep{2018MNRAS.477.1822M,2019MNRAS.488.3143B}, and to those predicted in \tng{} and \eagle{}. In \lgal{} the scatter with MRII resolution is slightly higher than those in \tng{} and \eagle{}. However, it is important to note that selecting complete samples based on halo mass observationally is very challenging. For instance, in \citet{2009ApJ...695..900Y}, the group mass may be underestimated or overestimated, especially when the satellite member is small.

In the next section, we further investigate the relation between the scatter in SHMR and general halo/galaxy properties, such as halo concentration, formation time, merger history, the mass of the black hole ($M_{\rm BH}$), the mass of cold gas ($M_{\rm coldgas}$), the star formation rate (SFR), and the mass-weighted age. We particularly focus on the effect of AGN feedback on the SHMR in simulations.

\section{Results}
\label{sec:results}

To enhance the precision of estimating the impact of various properties, we employ the residual $\Delta M_{*}$ - halo mass relation instead of the traditional SHMR. This is achieved by fitting the SHMR in log-log space with three parameters for each simulation:
 \begin{equation}
     {\rm log_{10}}(M_{*}/\Msun) = \alpha - e^{\beta {\rm log_{10}}(M_{h}/\Msun)+\gamma}.
     \label{eq:firstfit}
 \end{equation}
Recognizing that low-mass galaxies predominate over massive ones, we opt to stratify our samples based on the mass of their host haloes, employing a bin size of 0.1 dex. Bins containing fewer than 10 galaxies are excluded. Subsequently, we utilize the median stellar mass and halo mass within each bin to fit Eq.\ref{eq:firstfit}. The resulting fitting parameters are enumerated in Table~\ref{tab:firstfitting}. $\Delta M_*$ is defined as the deviation of the stellar mass from the fitted value at any given halo mass.
\begin{equation}
     \Delta M_* = {\rm log_{10}}(M_{*}/\Msun) - {\rm log_{10}}(M_{*, \rm fit}/\Msun).
     \label{eq:delta1}
 \end{equation}
Consequently, $\Delta M_*$ signifies the deviation around the median value of SHMR. Within each halo mass bin, we compute the standard deviation of $\Delta M_*$, denoted as $\sigma (\Delta M_{*}(M_{\rm h}))$ (hereafter referred to as $\sigma (\Delta M_{*, \rm h})$), which represents the scatter of the SHMR.

We also try to quantitatively assess the contributions of each property. Within each halo mass bin, we establish the relationship between $\Delta M_*$ and halo/galaxy properties as follows:
 \begin{equation}
     \Delta M_{*, \rm h} = a + b * {\rm log}_{10}(\rm Prop),
     \label{eq:secondfit}
 \end{equation}
where Prop refers to halo or galaxy properties, such as halo concentration, formation time, star formation rate, etc. It is noteworthy that the results obtained through linear fitting are comparable to those obtained using higher-order polynomial fittings. The residual scatter remaining after considering secondary properties is expressed as:
 \begin{equation}
     \Delta M_{*,\rm prop}(M_{\rm h}) = a + b * {\rm log}_{10}({\rm Prop}) - \Delta M_{*,\rm h}.
     \label{eq:delta2}
 \end{equation}
 The standard deviation of $\Delta M_{*, \rm prop}$ is then presented as $\sigma (\Delta M_{*, \rm prop})$, providing a quantitative measure of the scatter after accounting for the influence of the secondary property. Smaller $\sigma (\Delta M_{*, \rm prop})$ indicates a higher dependence on this property.

\begin{table}
	\centering
	\caption{Fitting parameters for the stellar mass - halo mass relation (\refeq{eq:firstfit}) for different simulations.}
	\label{tab:firstfitting}
	\begin{tabular}{lccr} 
		\hline
              & $\alpha$ & $\beta$ & $\gamma$ \\
		\hline
		\lgal & 11.63 & -0.76 & 9.53\\
		\lgalwo & 13.92 & -0.41 & 6.22\\
		\eagle & 12.34 & -0.56 & 7.42\\
            \tng & 12.16 & -0.69 & 8.84\\
		\hline
	\end{tabular}

\end{table}

In each subsection, we initially examine whether $\Delta M_{*}$ is an increasing, decreasing function, or independent of the properties. We subsequently measure the effectiveness of these properties in reducing $\sigma (\Delta M_{*})$.

\begin{figure*}
\centering
\subfloat{\includegraphics[width=1.6\columnwidth]{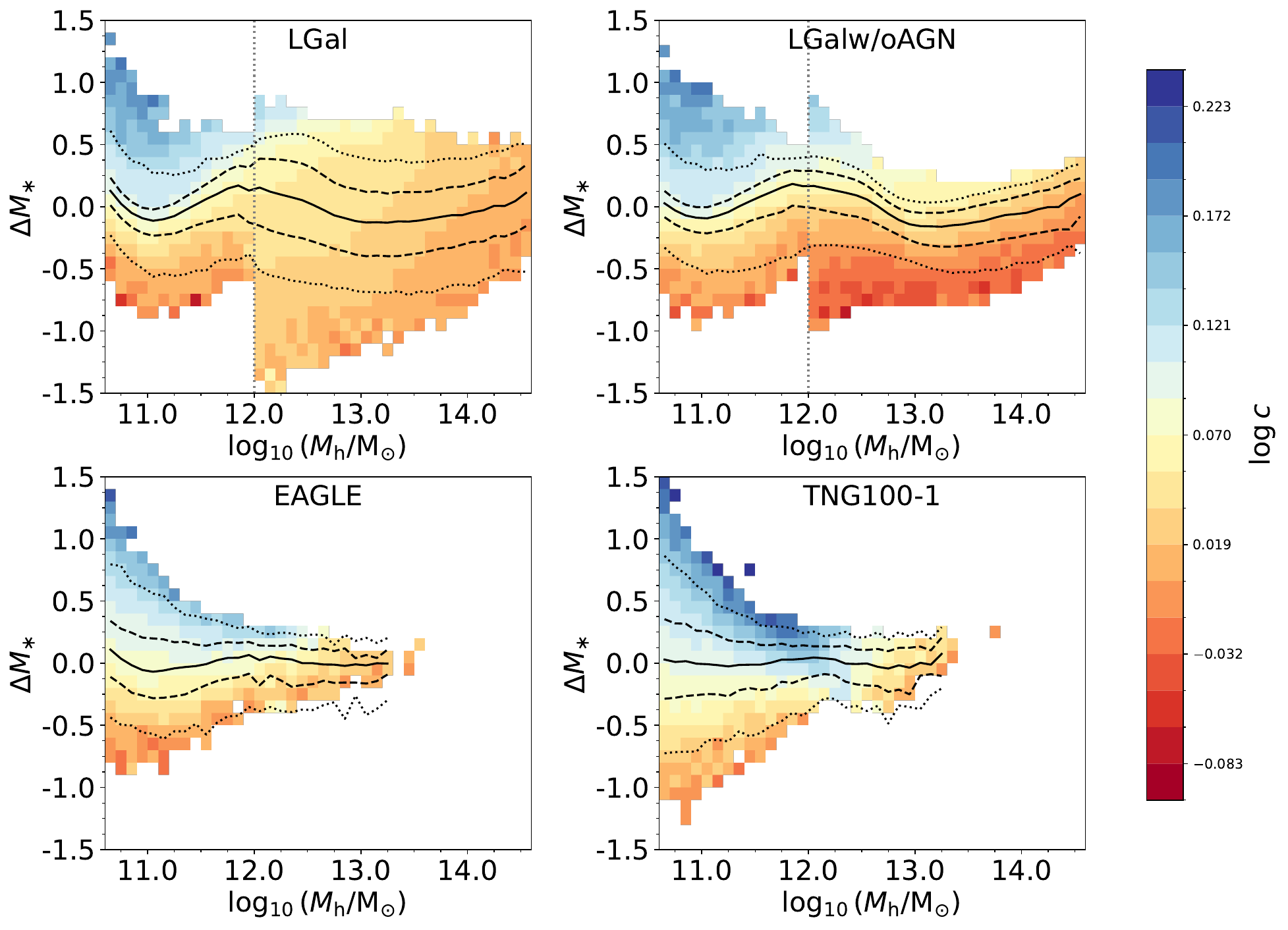}}\\
\subfloat{\includegraphics[width=1.6\columnwidth]{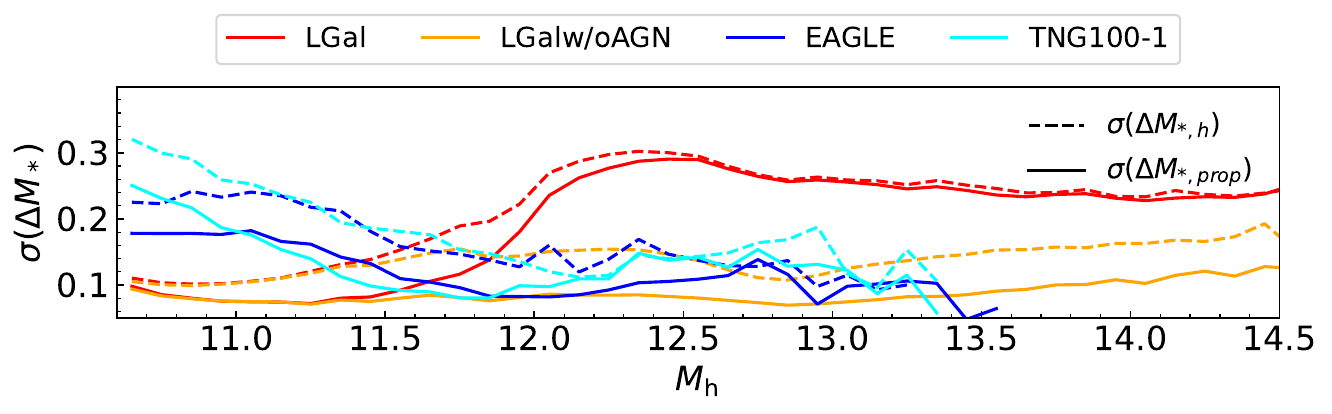}}
\caption{The distribution of delta stellar mass for central galaxies as a function of halo mass in \lgal, \lgalwo, \eagle, and \tng{} simulations. Black solid lines correspond to the medians of the distributions, while the black dashed (dotted) lines show the 16th (2nd) and 84th (98th) percentiles. The colours indicate the concentration, $c$, of haloes in each halo and delta stellar mass bin. The gray vertical dotted lines in the top panels represent the transition mass,  $M_{\rm h} = 10^{12}\Msun$, between MRII (left) and MR(right).} Bottom Panel: The standard deviation of $\Delta M_{*}$ as a function of halo mass. Different colours represent different simulations. Dashed lines show $\sigma(\Delta \rm M_{*, h})$ from SHMR while solid lines show $\sigma (\Delta \rm M_{*, prop})$ from the second fitting which considers halo concentration. See the text for details.
\label{fig:c}
\end{figure*}

\begin{figure*}
\centering
\subfloat{\includegraphics[width=1.6\columnwidth]{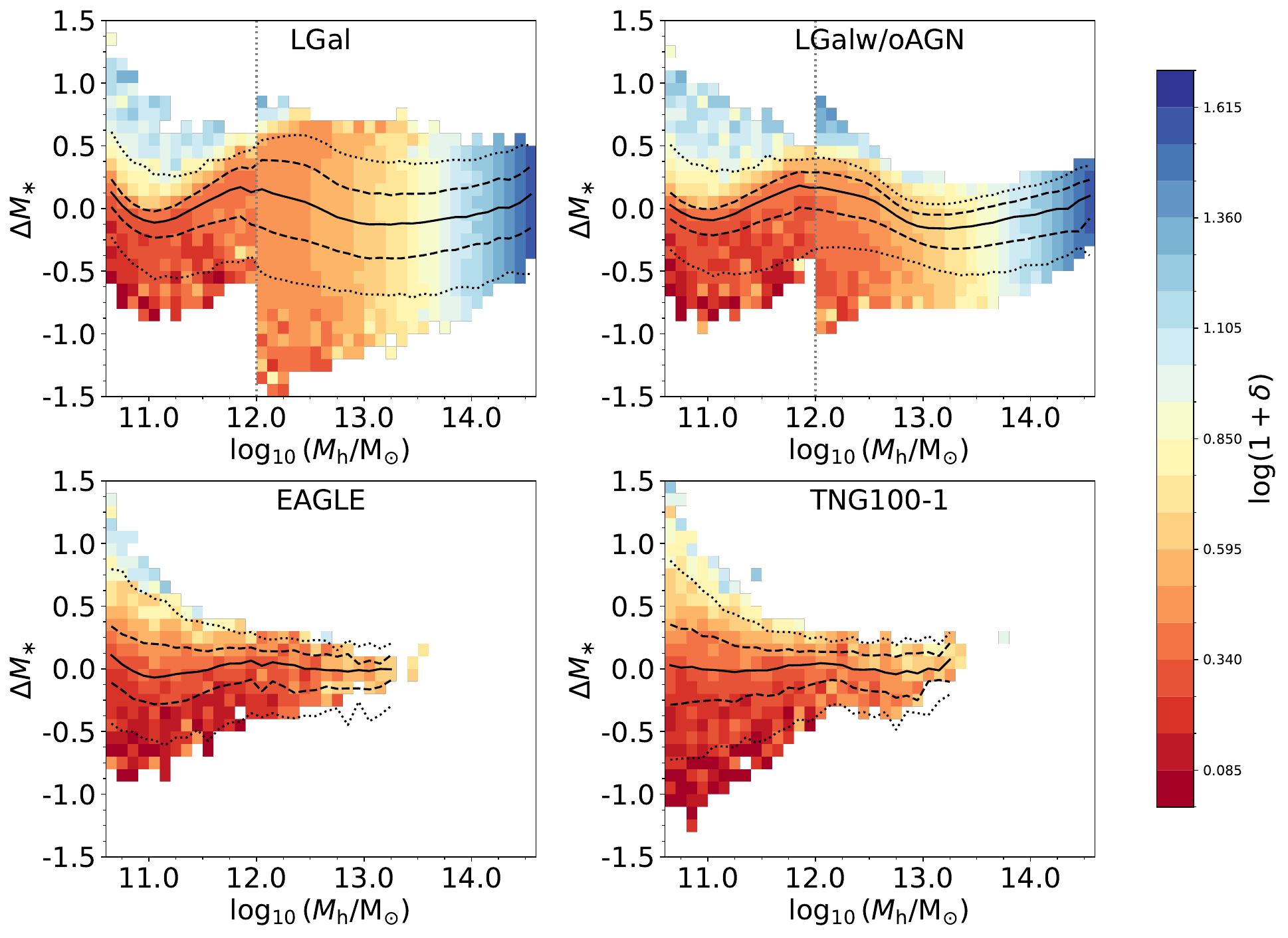}}\\
\subfloat{\includegraphics[width=1.6\columnwidth]{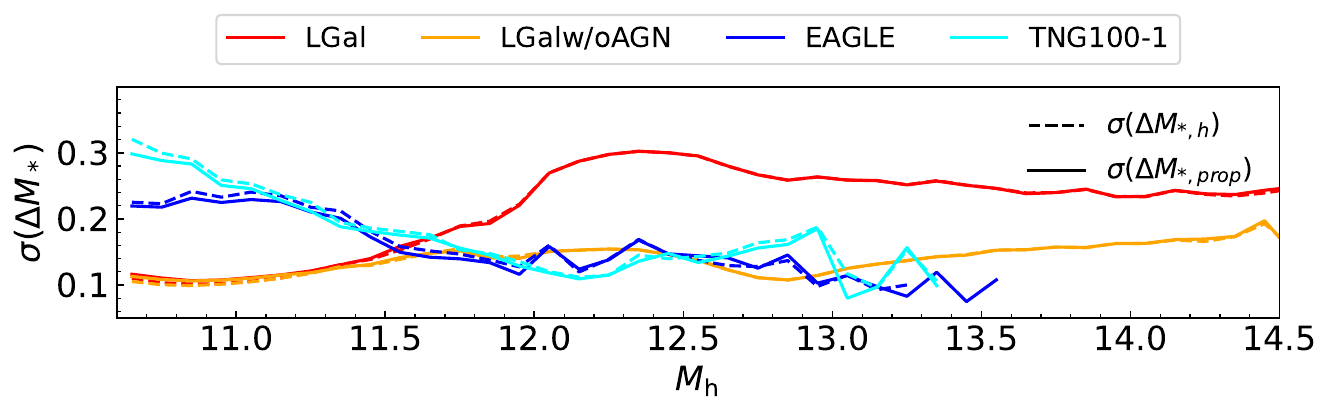}}
\caption{Same as Figure~\ref{fig:c} but each bin is coloured by the mean environmental overdensity, $\delta$, of the haloes. Solid lines in the bottom panel show $\sigma (\Delta \rm M_{*, prop})$ from the second fitting which considers $\delta$.}
\label{fig:density}
\end{figure*}

\begin{figure*}
\centering
\subfloat{\includegraphics[width=1.6\columnwidth]{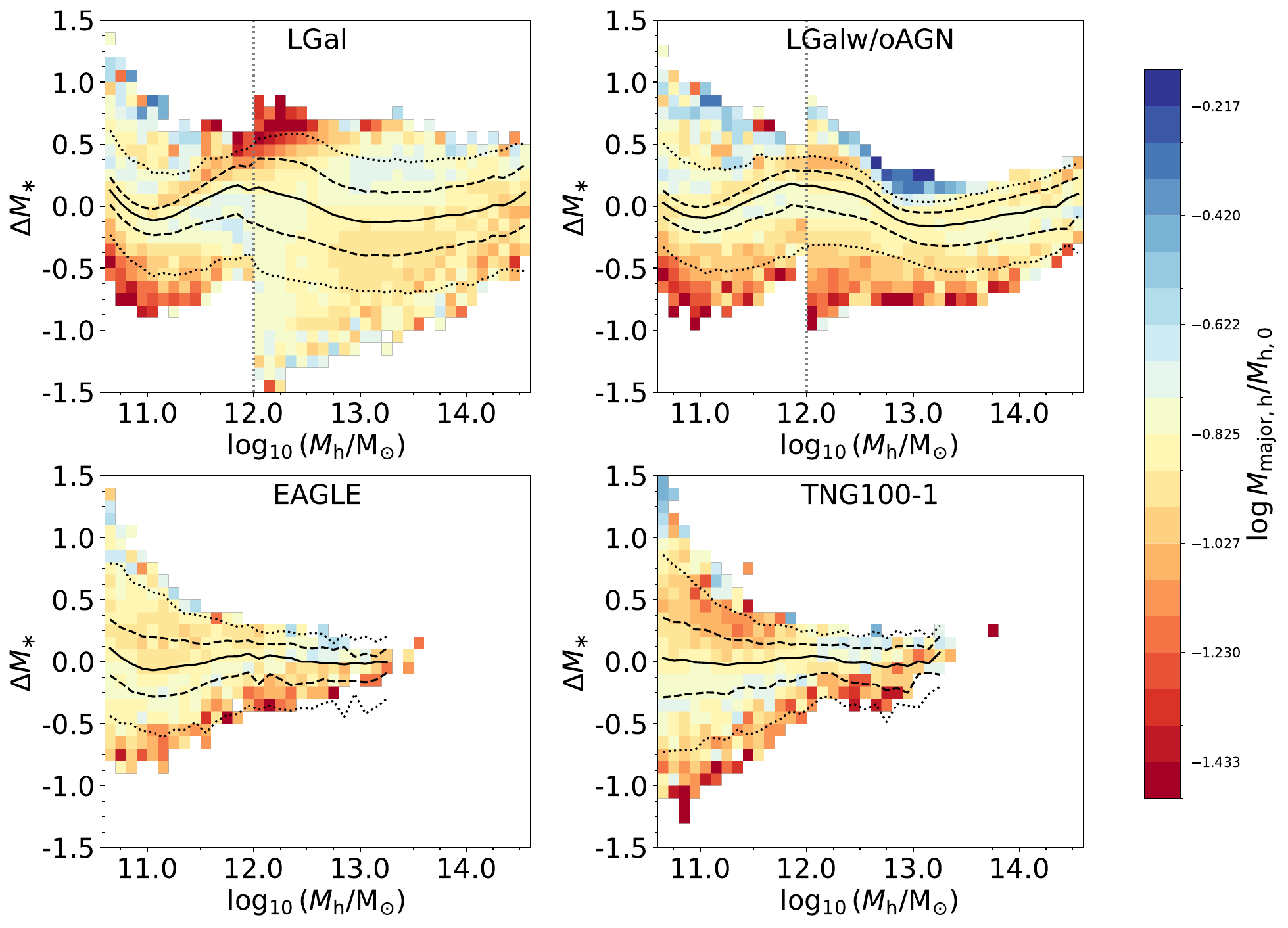}}\\
\subfloat{\includegraphics[width=1.6\columnwidth]{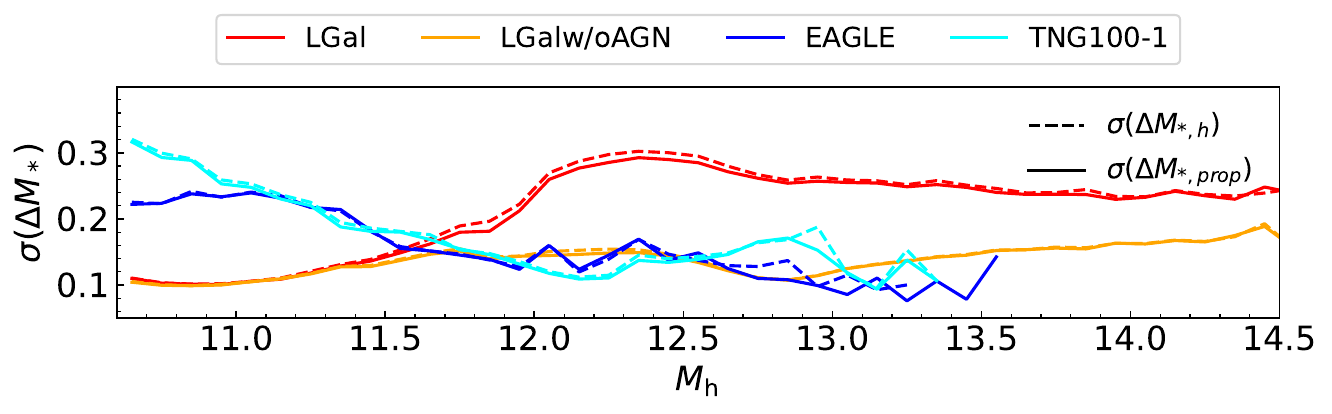}}
\caption{Same as Figure~\ref{fig:c} but each bin is coloured by the mean major-merged satellite mass fraction, $M_{\rm major}/M_{\rm h}$. Solid lines in the bottom panel show $\sigma (\Delta \rm M_{*, prop})$ from the second fitting which considers $M_{\rm major}/M_{\rm h}$.}
\label{fig:mm}
\end{figure*}

\subsection{Dependence on halo properties}
\label{sec:halo}
In this subsection, we investigate the impact of different halo properties on the scatter around the SHMR. All parameters depend on the total mass, total mass distribution and evolution within R$_{200c}$, if not stated otherwise.

\subsubsection{Halo concentration and formation time}
\label{sec:c}
The most heavily explored quantities as the candidates of secondary dependence of the SHMR include halo concentration and formation time \citep[e.g. ][]{2013MNRAS.435.1313H,2015MNRAS.452.1958H,2015ApJ...799..130R,2016MNRAS.460.1457S,2017MNRAS.465.2381M}. We compare the impact of halo concentration and formation time using consistent definitions across various simulations. Following the method in \citet{2012MNRAS.423.3018P}, we refer to $c={v_{\rm max}}/{v_{\rm vir}}$ as the concentration of the halo, where $v_{\rm max}$ is the maximum rotational velocity of the halo and $v_{\rm vir}$ is the virial velocity.

\reffig{fig:c} shows the $\Delta M_{*}$ - halo mass relation for the four simulations. 
We categorize galaxies based on their halo masses using a bin size of 0.1 dex. Within each halo mass bin, we further divide their delta stellar mass into a bin size of 0.1 dex, and then the cells are colour-coded based on the average concentration of the galaxies in this cell, as illustrated by the colour bar. 
This approach effectively compresses the dynamical range of concentration and highlights the dependence on concentration in each halo mass bin. The solid/dashed/dotted lines in each panel represent the 50\% / 16\%(84\%) / 2\%(98\%) values of the $\Delta M_*$ - $M_{\rm h}$ relation. For a fixed halo mass, highly concentrated haloes tend to host more massive galaxies than haloes with low concentrations. This trend is relatively strong at low masses in all the simulations and is weak at high masses except for \lgalwo{}. 

The formation time ($z_{\rm f}$) is also found crucial in determining galaxy properties. Indeed, previous studies \citep{2004MNRAS.355..819G, 2002ApJ...568...52W, 2009ApJ...707..354Z, 2011MNRAS.415L..69J, 2014MNRAS.441..378L,2015MNRAS.450.1521C} have demonstrated a robust correlation between halo formation time and concentration, where higher concentration corresponds to earlier formation time. We find a similar overall trend on $z_{\rm f}$ as concentration, as shown in the Appendix Figure~\ref{fig:tf}, that early-formed haloes host more massive galaxies at fixed halo mass in all simulations and across all halo mass range.

The reliance on halo concentrations and formation time is based on physical reasons: i) host haloes that form earlier will have more time to gather gas and create stars; and ii) highly concentrated haloes have steeper potential wells that can retain more material against feedback, potentially increasing thermal and kinetic energies. 

At low masses, similar dependencies on halo concentration and formation time have been reported in many previous works, utilizing semi-analytical models \citep[e.g.][]{2013MNRAS.431..600W, 2017MNRAS.470.3720T, 2018ApJ...853...84Z} or hydrodynamical simulations \citep[e.g.][]{2017MNRAS.465.2381M, 2018MNRAS.480.3978A, 2020MNRAS.492.2739X, 2020MNRAS.496.1182M,2021NatAs...5.1069C}, indicating that haloes which are highly concentrated and formed earlier tend to host more massive galaxies. However, the dependency at high masses is still a topic of debate. \citet{2021MNRAS.505.5117Z} found that for galaxy clusters with $M_{\rm h}\sim 10^{14.24}\Msun$, samples with higher stellar mass also have higher halo concentration.
In our study, at high masses, the dependency on halo concentration and formation time is overshadowed by AGN feedback. However, \cite{2020MNRAS.493..337B} reported that the impact of formation time continues to rise, being most significant at highest halo masses in empirical models like the UniverseMachine \citep[UM,][]{2019MNRAS.488.3143B}. All detailed differences rely on the different treatments of baryonic processes employed in various models and simulations.

We further show the standard deviation of $\Delta M_{*}$ as function of halo mass in the bottom panel of Figure~\ref{fig:c}. Lines with different colours represent different simulations. The dashed lines show the standard deviation of $\Delta M_{*,\rm h}$ from Eq.~\ref{eq:delta1}, which only adopts halo mass to fit stellar mass. The solid lines show the standard deviation of $\Delta M_{*,\rm prop}$ from Eq.~\ref{eq:delta2}, which takes both halo mass and secondary properties, concentration, into consideration. The greater the difference between solid and dashed lines, the stronger the dependence on this property.
We observe that removing the dependency on concentration could significantly decrease scatter around the SMHR. At low masses, around $M_{\rm h} \sim 10^{11}\Msun$, the scatter could potentially be decreased by about 40\%. At higher masses, the effect varies among simulations, with more pronounced impacts seen in \eagle{} and \tng{} compared to \lgal{}. This aligns with findings in \citet{2017MNRAS.465.2381M} which report a stronger correlation with halo concentration when utilizing simulations incorporating full baryonic processes. This is because including baryonic processes tends to steepen the density profile \citep{2015MNRAS.451.1247S}, resulting in higher concentrations compared to DMO simulations. This amplifies the reliance on halo properties in hydrodynamical simulations. 

Interestingly, we observe a notable difference of the halo concentration dependence at high masses, $M_{\rm h}>10^{12}\Msun$ in simulations with and without AGN feedback. \lgalwo{} shows a stronger dependence on concentration at high masses, similar to that at low masses, while other simulations show a much weaker dependence at high masses. Detailed comparisons in the bottom panel of Figure~\ref{fig:c} show that for \lgalwo{}, scatters are reduced by 40\% at high masses, much higher than those in \eagle{}, \tng{}, and \lgal{}. Further, it suggests that the relatively strong effect of halo concentration on scatters around the SHMR is probably attributed to the less effective AGN feedback adopted in \eagle{} and \tng{}. This is because, in simulations with AGN feedback, the feedback effect is a highly nonlinear process that effectively modifies star formation activities in massive haloes. Some galaxies experiencing strong AGN feedback are quenched and cease further star formation, leading to a constant stellar mass since the quenching event. However, the host haloes of these galaxies continue to grow, causing the stellar mass to decouple from the halo mass and other halo properties. In \lgalwo{}, where there is no AGN feedback, all galaxies naturally grow following the halo properties, resulting in a stronger dependence on halo properties and a smaller scatter compared to \lgal{} at high masses. Such dilutions are also observed in the dependence on other halo properties.

\subsubsection{Environmental overdensity}
The influence of large-scale environments on galaxy formation is widely acknowledged \citep[e.g.][]{2010ApJ...721..193P,2019MNRAS.485..464L,2020MNRAS.498.1839X}. While the environmental effect may be linked to halo concentration and formation time, we do not intend to distinguish their relative impacts, as measuring the environments is easier than the latter two. To explore its impact on the SHMR, we define $\delta = \rho/\Bar{\rho} - 1$ as the local overdensity for each halo in the simulations, where $\rho$ represents the local density and $\Bar{\rho}$ is the mean density of the Universe. In practice, we initially partition the simulation box into cubes with a side length of $1\mpch$ and calculate the mean density for each cubic cell. Subsequently, following the \textsc{cloud-in-cell} method \citep{2016MNRAS.460.3624S}, the overdensity of each halo at a given position is computed by using the surrounding cells with a Gaussian smoothing kernel that has a half-height full width of $2.1\mpch$. We have investigated cell sizes of 0.5\mpch, 1\mpch, 2\mpch, and 5\mpch, and our primary findings remain unaffected by this choice.

In Figure~\ref{fig:density}, we utilize a colour-coded scheme based on the average $\delta$ to depict the SHMR. The broad patterns resemble those observed for halo concentration at low masses. A discernible trend is apparent across all simulations, indicating a tendency for massive galaxies to reside in regions of higher density at a fixed halo mass. At higher masses, all galaxies tend to inhabit dense regions, exhibiting a little trend at a fixed halo mass. This finding suggests that a denser environment can promote star formation slightly, with the influence of the environment saturating at approximately $10^{12}\Msun$. At higher masses, the local environment is sufficiently dense so that a higher overdensity contributes minimally to the stellar mass. 

Previous studies have reached similar conclusions at low masses, while at high masses this dependence varies \citep{2017MNRAS.470.3720T, 2018MNRAS.480.3978A, 2018ApJ...853...84Z, 2021MNRAS.507.5320Z}. Utilizing data from the GAMA survey \citep{2015MNRAS.452.2087L}, \citet{2017MNRAS.470.3720T} found that at low masses, galaxies in voids have a lower stellar-to-halo mass ratio than those in knots, while this trend reverses at high masses. \citet{2021MNRAS.507.5320Z} also find this reverses trend by combing SDSS DR7 data with ELUCID simulation. \citet{2024arXiv240207995W} report that using the information about the local environment will significantly increase the accuracy for predicting stellar mass with machine learning in TNG300. 

When AGN feedback is removed, as illustrated for \lgalwo{}, this positive relationship with environments remains up to higher halo mass levels of $10^{13}\Msun$, stronger than those in other simulations at high masses. 
The impact of AGN on environmental dependence exhibits a resemblance to that of concentration/formation time, indicating that the inclusion of AGN breaks the link between the growth of stellar components and halo accretion.

A detailed examination of the dependence on environments is presented in the bottom panel of Figure~\ref{fig:density}. It shows that even at the lowest masses, the scatter is only reduced by 5\%, while at high masses the influence of density almost vanishes. The slight effect, compared to the visible distinctions as indicated by the coloured pixels in the upper panels, is because within one sigma ranges, the density does not significantly contribute to the 1$\sigma$ scatter although it becomes more noticeable at a 2 $\sigma$ level. Such weaker dependence could be further reduced as demonstrated by \citet{2017MNRAS.465.2381M} that the dependence on the environment diminishes when considering the role of concentration/formation time. In conclusion, the reliance on environments is notably weaker compared to that on halo concentration and halo formation time. 

\begin{figure*}
\centering
\subfloat{\includegraphics[width=1.6\columnwidth]{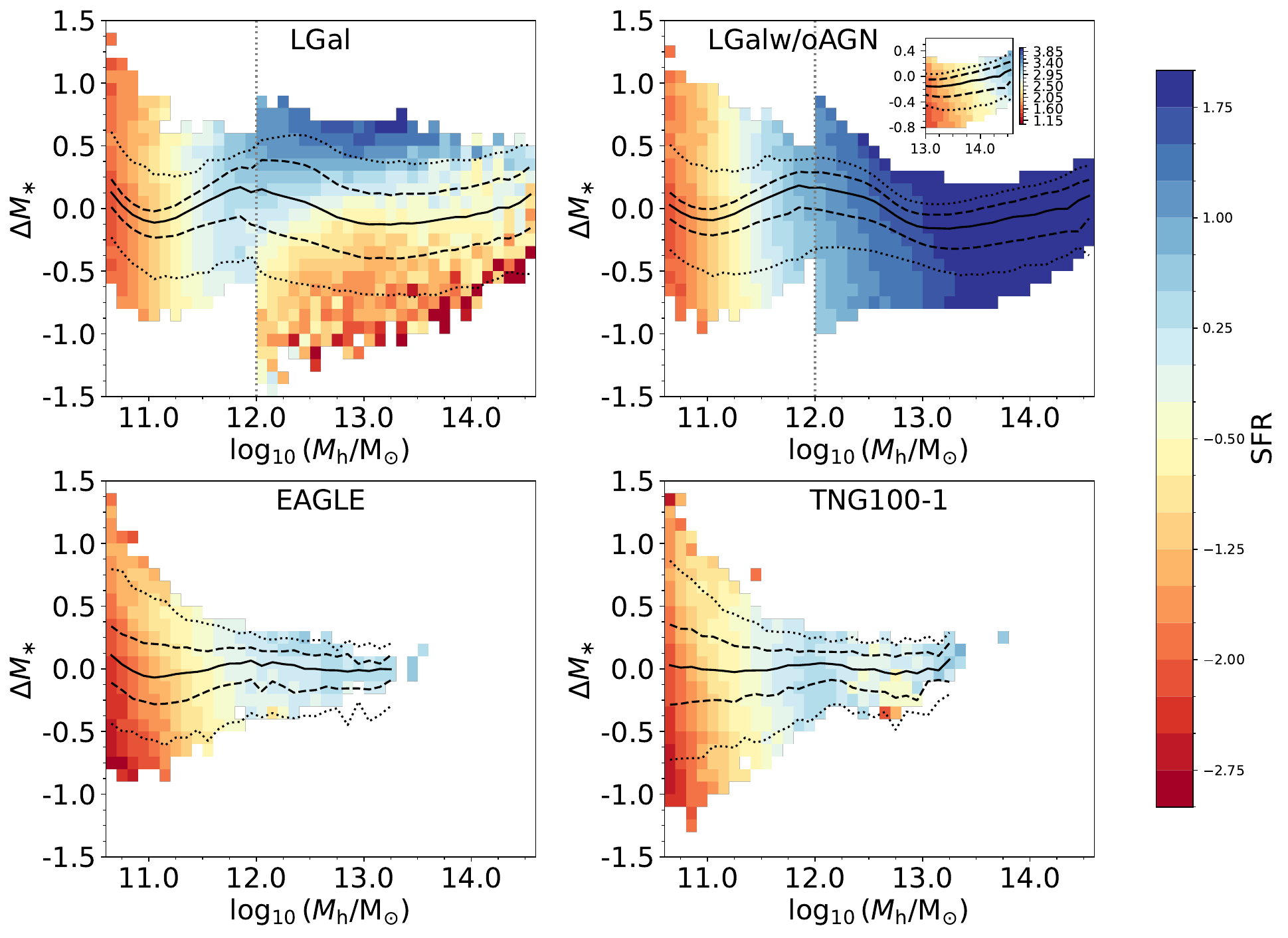}}\\
\subfloat{\includegraphics[width=1.6\columnwidth]{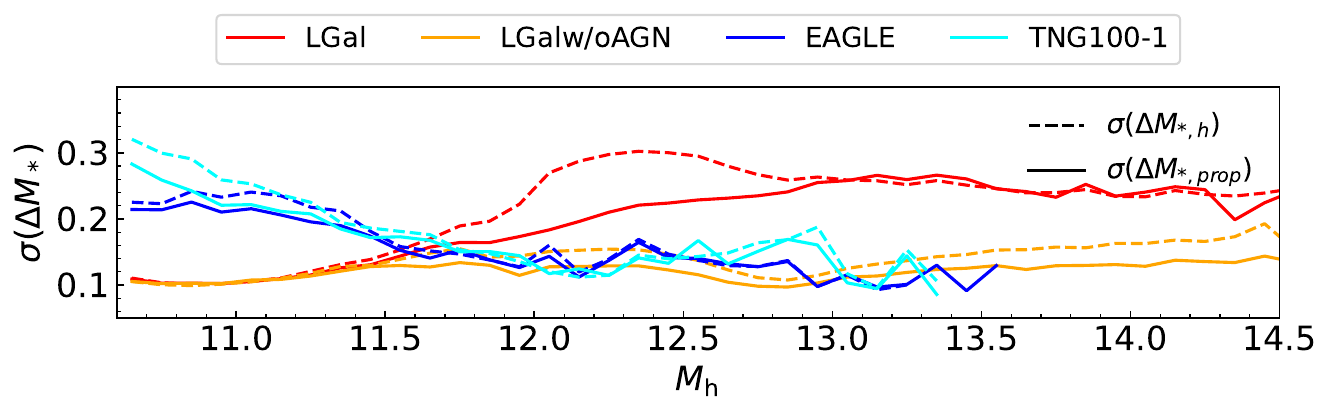}}
\caption{Same as Figure~\ref{fig:c} but each bin is coloured by the average SFR of central galaxies at $z=0$. Solid lines in the bottom panel show $\sigma (\Delta \rm M_{*, prop})$ from the second fitting which considers SFR. The subplot in \lgalwo{} highlights the differences at high masses with a different colour-bar range.} 
\label{fig:sfr}
\end{figure*}

\subsubsection{Halo major merger}

Halo mergers play an important role in galaxy formation and evolution, particularly when a galaxy devours a massive satellite. Here, we focus on haloes that have undergone major mergers throughout their entire assembly history, where a major merger event is defined as when a satellite falls into a host halo with a mass ratio exceeding 1/3. We denote $M_{\rm major}$ as the sum of the infall mass of all major-merged satellites for each host. 

In Figure~\ref{fig:mm}, the scatter of the SHMR is colour-coded by their average $M_{\rm major}/M_{\rm h,0}$ in each cell, where $M_{\rm major}/M_{\rm h,0}$ represents the total major merger halo mass normalized to the host halo mass at z$\sim$0. The influence of major mergers appears to be relatively weak if any. In general, most haloes tend to have a similar fraction of major merger halo mass at approximately 10\% to 20\% of the total halo mass at redshift 0 despite the halo mass range in all the simulations. 
Previous studies \citep[e.g.,][]{2008MNRAS.384....2G,2010ApJ...725.2312O,2016MNRAS.458.2371R,2017MNRAS.464.1659Q,2018MNRAS.475..648P,2018MNRAS.477.1822M,2020MNRAS.493..337B} find that mergers have much less impact on galaxy growth at low masses compared to star formation. Therefore, a clear dependence on mergers is not anticipated. At low masses, Figure~\ref{fig:mm} shows a slightly negative correlation within the 1$\sigma$ region, which might reverse for outliers in \eagle{} and \tng{}. In \lgal{}, there is a positive relationship between stellar mass and accreted mass fraction below $10^{10}\Msun$ within 1$\sigma$ ranges, but this changes at higher masses. The trends in the outskirts are more intricate. At high masses, there is a weak positive association between $M_{\rm major}/M_{\rm h,0}$ and scatter in all models. This is expected because mergers could play a significant role in the growth of galaxies at these masses in simulations and models \citep{2008MNRAS.384....2G,2017MNRAS.464.1659Q,2018MNRAS.475..648P,2018MNRAS.477.1822M,2020MNRAS.493..337B}.

The quantitative impact of $M_{\rm major}$ is presented in the bottom panel of Figure~\ref{fig:mm}. Overall, halo major mergers do not significantly affect the scatter in the SHMR at low masses for all simulations. The largest difference in scatter is only 5\% for \lgal{} in the mass range $10^{12}<M_{\rm h}<10^{13}\Msun$.

In addition to $M_{\rm major}/M_{\rm h,0}$, the last major merger redshift, $z_{\rm major}$ could be relevant.  $z_{\rm major}$ represents the redshift when it underwent its last major merger. A larger $z_{\rm major}$ indicates that the galaxy underwent its last major merger earlier and has had a fairly steady growth history since then. This could be linked to the halo formation time.  Figure~\ref{fig:lmt} demonstrates that at lower masses, as anticipated, massive galaxies exhibit a higher $z_{\rm major}$. The influence of $z_{\rm major}$ weakens significantly at high masses due to AGN feedback. Detailed discussion about the influence of $z_{\rm major}$ is presented in Appendix~\ref{app:lmt}.

\begin{figure*}
\centering
\subfloat{\includegraphics[width=1.6\columnwidth]{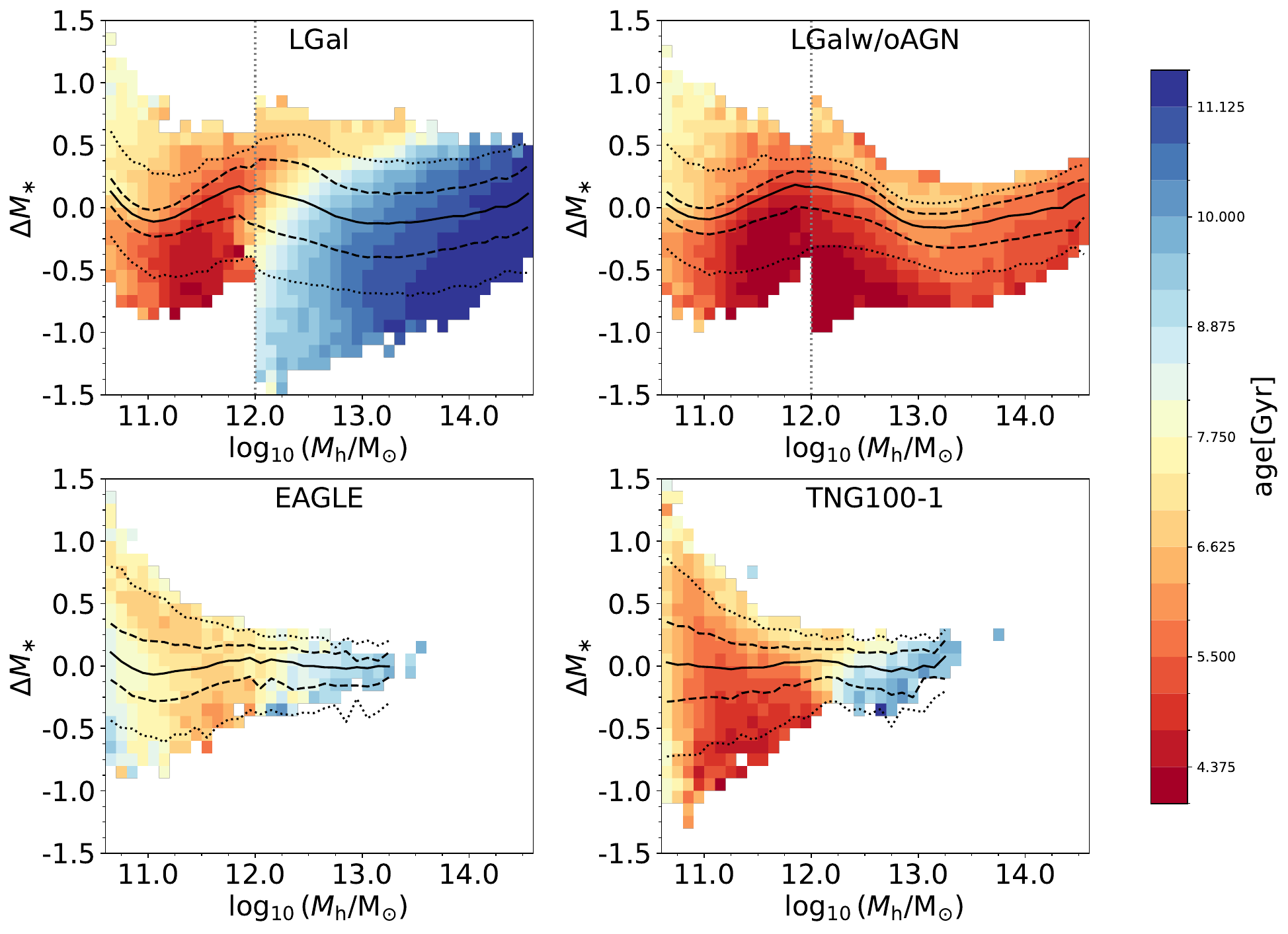}}\\
\subfloat{\includegraphics[width=1.6\columnwidth]{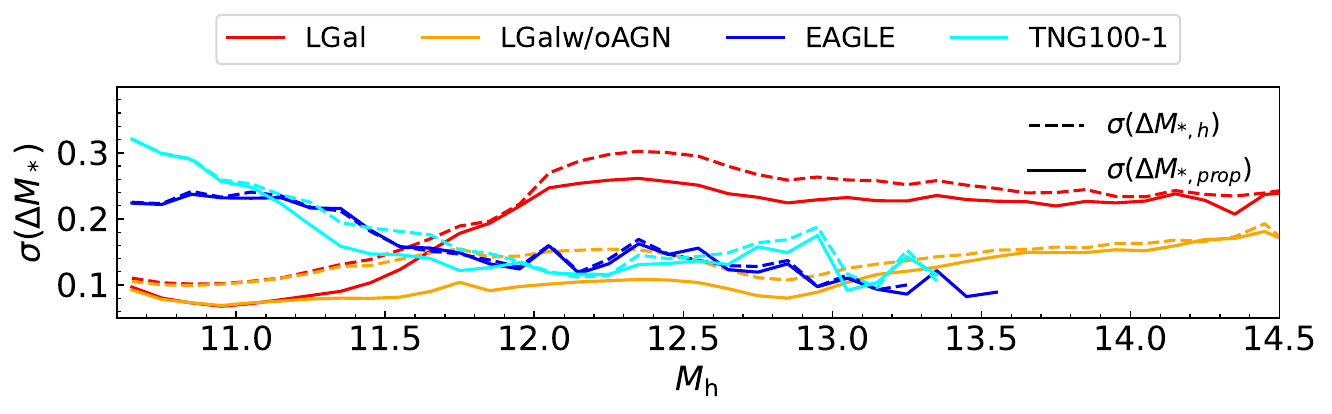}}
\caption{Same as Figure~\ref{fig:c} but each bin is coloured by the mass-weighted age of the central galaxies. Solid lines in the bottom panel show $\sigma (\Delta \rm M_{*, prop})$ from the second fitting which considers age.}
\label{fig:age}   
\end{figure*}

\subsection{Dependence on galaxy properties}
\label{sec:galaxy}

Star formation is not solely determined by the potential of the host dark matter halo, but also depends on more complex baryonic physics closely related to halo formation history and interactions between dark matter and baryons. 
In this section, we investigate how the SFR, cold gas mass ($M_{\rm coldgas}$), black hole mass ($M_{\rm BH}$) and mass-weighted age contribute to the scatter in the SHMR.

\subsubsection{Star formation rate}
\label{sec:sfr}

The star formation rate is closely linked to the stellar components within galaxies. Figure~\ref{fig:sfr} depicts the SHMR colour-coded by the SFR. In \lgal{} and \lgalwo{}, a positive correlation between SFR and stellar mass is observed within a specific halo mass range across all halo mass categories. At higher masses, this relationship is enhanced by AGN feedback when comparing \lgal{} and \lgalwo{}. A similar positive correlation between SFR and stellar mass is also found in the \eagle{} and \tng{} simulations for low mass systems. However, this trend weakens significantly for $M_{\rm h}>10^{12}M_{\odot}$, indicating a complex interplay with halo accretion history and baryonic processes. A mildly positive relationship with SFR is identified at $M_{\rm h}\sim 10^{12.8}\Msun$ in \tng{}, possibly influenced by the AGN feedback model. \citet{2023ApJ...959....5L} reports that at fixed halo mass, star-forming galaxies have larger stellar mass than passive galaxies in \lgalaxies{}. \citet{2023MNRAS.522.3188W} also finds that star-forming galaxies are more massive at fixed halo mass based on stellar mass from \citet{2003ApJS..149..289B} and halo mass from \citet{2007ApJ...671..153Y}. These are consistent with what we found. Nevertheless, a recent study by \citet{2024NatAs.tmp...42S} suggests that, at fixed halo mass, massive galaxies exhibit lower SFR compared to smaller galaxies in the CALIFA survey \citep{2014A&A...569A...1W}.

The bottom panel of Figure~\ref{fig:sfr} shows the quantitative impact of SFR. The correlation between SFR and the scatter around the SHMR varies with halo masses. At $M_{\rm h}\sim 10^{12.2}\Msun$, the scatter significantly decreases when incorporating SFR in \lgal{}, indicating a strong dependence of SFR on $\Delta M_{*}$. This dependency weakens as we move towards both higher and lower halo masses and eventually disappears at the extreme mass ends in \lgal{}. At the very massive ends, AGN feedback predominates in nearly all cases, mitigating the impact of SFR. Conversely, in both \eagle{} and \tng{}, a clear correlation with SFR is noticeable at lower masses and disappears for haloes exceeding $\sim 10^{11.5} \Msun$. The different dependence on SFR may be attributed to discrepancies in the treatment of star formation and AGN feedback among different simulations. Notably, in \lgalwo{}, the SFR continues to be strongly associated with the scatter around the SHMR towards the very massive end without AGN feedback.

\begin{figure*}
\centering
\subfloat{\includegraphics[width=1.6\columnwidth]{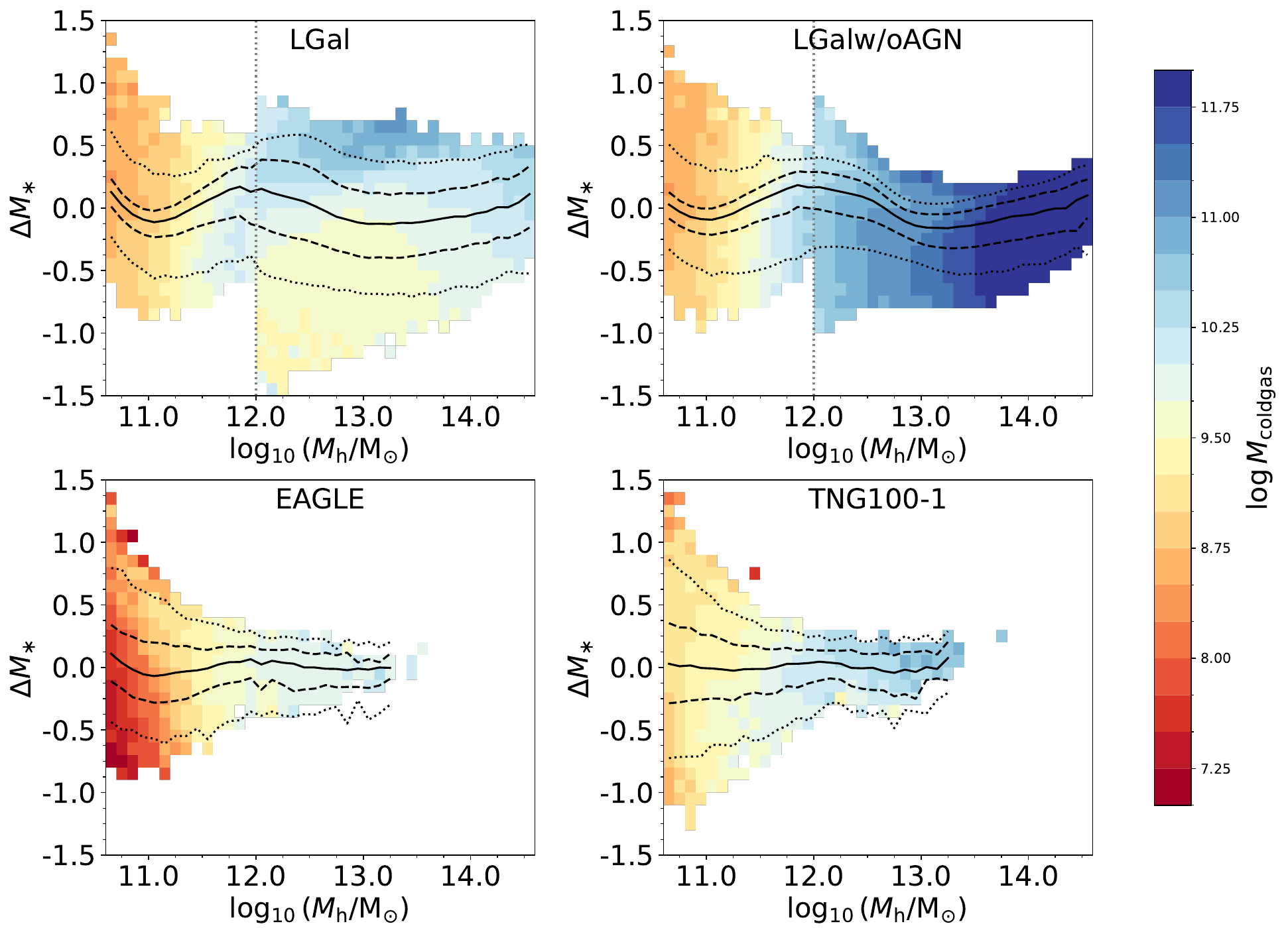}}\\
\subfloat{\includegraphics[width=1.6\columnwidth]{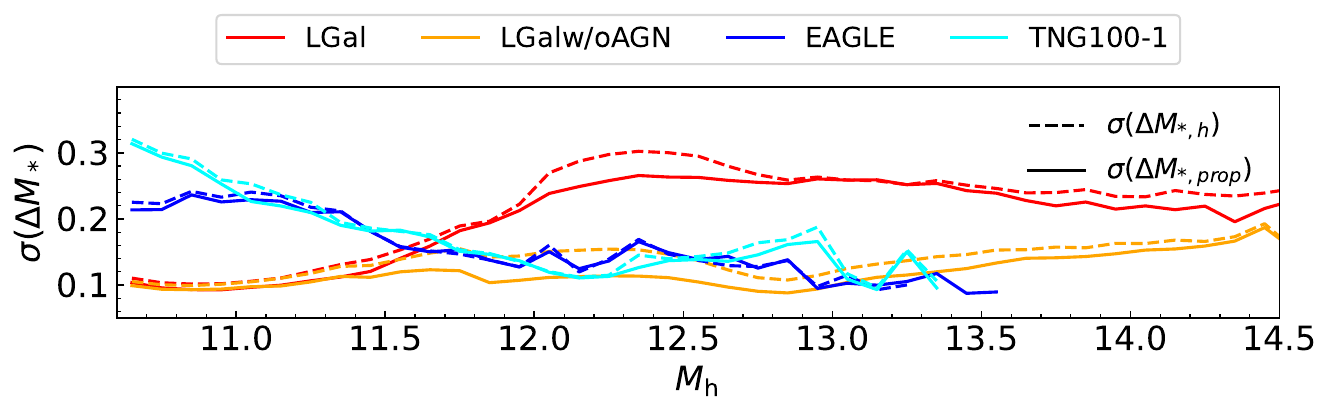}}
\caption{Same as Figure~\ref{fig:c} but each bin is coloured by the average cold gas mass of the central galaxies. Solid lines in the bottom panel show $\sigma (\Delta \rm M_{*, prop})$ from the second fitting which considers cold gas mass.} 
\label{fig:cg}   
\end{figure*}

\begin{figure*}
\centering
\subfloat{\includegraphics[width=2\columnwidth]{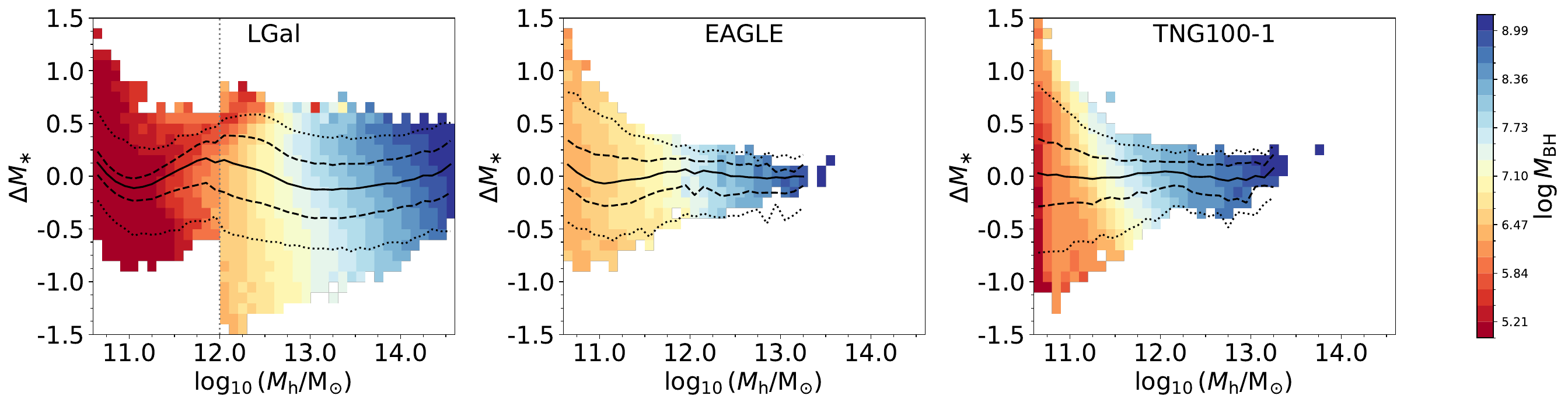}}\\
\subfloat{\includegraphics[width=1.6\columnwidth]{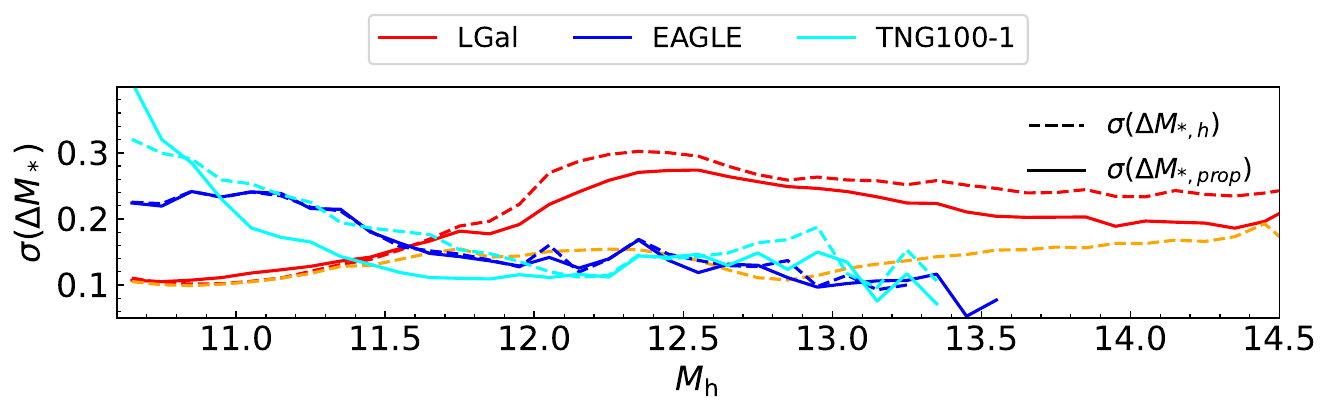}}
\caption{Same as Figure~\ref{fig:c} but each bin is coloured by the average mass of the supermassive black holes in \lgal{}, \eagle{}, \tng{}. Solid lines in the bottom panel show $\sigma (\Delta \rm M_{*, prop})$ from the second fitting which considers black hole mass.}
\label{fig:bh}   
\end{figure*}

\subsubsection{Age}
\label{sec:age}

A closely related galaxy property to the whole star formation history is the mass-weighted stellar age. Figure~\ref{fig:age} depicts the SHMR colour-coded by age. 
At low masses, all simulations indicate a positive correlation between age and stellar mass that older galaxies are more massive. Galaxies formed earlier tend to have more time to accumulate stellar mass. Such dependence in \eagle{} is rather weak. On the contrary, at high masses, all simulations with full physics demonstrate an inverse relationship between stellar mass and age, indicating that more massive galaxies tend to be younger. This could be explained by that in this mass range, galaxies quenched earlier are older and less massive, while galaxies quenched later or remaining active are younger and more massive. When AGN feedback is deactivated and galaxies do not get quenched, as in \lgalwo{}, galaxy stellar mass growth in massive halos follows similar formation mechanisms as those in low mass halos. Therefore, older galaxies have had more time to form stars, leading to a consistent correlation between galaxy age and the scatter around the SHMR across all mass ranges.

The relationship between age and stellar mass is still far from conclusive in previous literature. Several studies \citep{2021MNRAS.502.4457G,2022A&A...664A..61S,2024NatAs.tmp...42S} have reported that galaxies with older stellar populations are more massive at fixed halo mass. However, \citet{2022ApJ...933...88O} found that low-mass galaxies are older at $M_{\rm h} < 10^{12}\Msun$, while high-mass galaxies are older at large halo masses, which is contrary to our \red{predictions}. \citet{2019MNRAS.482.3261K} found that age does not correspond to the scatter in the SHMR; instead, it is related to the scatter in the stellar mass-$V_{\rm max}$ relation.

The age dependence of the scatter around the SHMR is depicted in the bottom panel of Figure~\ref{fig:age}. 
In \lgal{}, including age leads to a significant decrease in scatter across all halo mass ranges except for haloes at $10^{12}\Msun$. This decrease amounts to up to 30\% at low masses and 15\% at high masses, indicating a tight correlation between age and scatter. In \lgalwo{}, age contributions decrease with increasing halo masses, down to approximately 5\% of the scatter at high masses. Age merely contributes to the scatter in \eagle{} but has a stronger effect in \tng{} by up to 20\% in \tng{}, indicating the strong influence of different sub-grid physics.

\subsubsection{Cold gas mass}
\label{sec:cg}

Another galaxy property of interest is cold gas. The star formation rate in all simulations is related to the cold gas mass, although different thresholds and formulas are adopted. Figure~\ref{fig:cg} depicts the SHMR colour-coded by $M_{\rm coldgas}$. It is important to note that the definition of cold gas differs among simulations, which may introduce additional differences in cold gas dependence. At low masses, almost all simulations predict no cold gas mass dependence for galaxies with fixed halo mass while \eagle{} shows a slightly positive trend. At high masses, galaxies below the median stellar mass exhibit significantly lower $M_{\rm coldgas}$ in \lgal{}, in line with those found by \citet{2021NatAs...5.1069C} in the SIMBA simulation. Galaxies with lower cold gas masses also exhibit lower SFR, as illustrated in Figure~\ref{fig:sfr}, suggesting a connection between the decline in SFR and cold gas caused by AGN feedback efficiency. In contrast, in \eagle{} and \tng{} no evident dependence of the scatter on cold gas is observed, which indicates the inefficient AGN feedback in these two simulations. This is further supported by the vanishing relation between the scatter and the cold gas in AGN feedback-free simulations as shown for \lgalwo{}.

The bottom panel of Figure~\ref{fig:cg} provides an assessment of the scatter in the SHMR as it relates to cold gas. The degree of scatter exhibits variability across different simulations. In the \lgal{} simulation, the scatter is significantly reduced, reaching up to 15\% at intermediate and high masses. Conversely, in the \eagle{} and \tng{} simulations, the reduction in scatter is less pronounced, amounting to less than 10\%, and is only observable at the extreme ends of the SHMR. These differences may stem from different star formation recipes, given the varying relationships between SFR and cold gas in each simulation. Additionally, it is important to note that we employ slightly different definitions of cold gas among these simulations, thereby introducing additional complexities.

\subsubsection{Black hole mass}
\label{sec:bh}

We have undertaken a similar analysis on the Supermassive Black Holes (SMBHs) in the simulations, considering the close relationship between SMBH mass and AGN feedback in all simulations (see Section~\ref{sec:data} for details). Figure~\ref{fig:bh} shows the SHMR colour-coded by $M_{\rm BH}$ in \lgal{}, \eagle{} and \tng{}. It does not include results from \lgalwo{} because $M_{\rm BH} = 0$ for all galaxies in this simulation. In \lgal{}, galaxies with more massive SMBHs usually have higher stellar masses at both low and high mass ranges, but this trend reverses at intermediate masses. In \eagle{} simulation, there is no significant trend between the scatter and SMBH mass, which is also reported by \citet{2020MNRAS.499.3578C}. A positive relationship is evident in \tng{} at lower masses, suggesting a co-evolution of the stellar components and the SMBH, which disappears at higher masses. The discrepancy among different simulations could be due to the different SMBH growth and feedback model in each simulation. This relationship suggests a complex co-evolution between galaxies and their central supermassive black holes \citep[see][for a review]{2013ARA&A..51..511K}. While such a dependence is expected at lower masses, its counter-intuitive manifestation at higher masses is noteworthy. In massive systems, potent feedback from AGN counteracts gas cooling, maintaining these galaxies in a quenched state. Stellar and SMBH mass growth predominantly stems from mergers and merger-related events. Previous studies indicate that at higher masses, mergers drive stellar mass growth in galaxies \citep{2008MNRAS.384....2G,2010ApJ...725.2312O,2016MNRAS.458.2371R,2017MNRAS.464.1659Q,2018MNRAS.475..648P,2018MNRAS.477.1822M,2020MNRAS.493..337B}, and both mergers and merger-induced gas inflow drive SMBH growth, thereby sustaining the co-growth dynamic between stellar mass and SMBH mass. However, at intermediate masses where AGN feedback initiates, the relationship between SMBH mass and stellar mass becomes more intricate. In certain systems, AGN feedback strongly suppresses star formation, while in others, it may not be sufficiently effective. Therefore, the association between stellar mass and SMBH mass in haloes of a given mass may depend on specific models of baryonic physics.

The bottom panel in Figure~\ref{fig:bh} elucidates the quantified relationship between the scatter in the SHMR and the SMBH mass in the simulations. This dependence closely mirrors that observed with cold gas in both the \lgal{} and \eagle{} simulations. It is worth noting that the dependence on SMBH is considerably more pronounced in the \tng{} simulation. Specifically, at intermediate masses, the dependence on SMBH can account for up to 40\% of the scatter, while at high masses, the SMBH contributes to approximately 20\% of the scatter. This indicates the substantial co-evolution scenario of SMBH and host galaxy, particularly in the \tng{} simulation.

\begin{figure*}
\centering
\includegraphics[width=1.6\columnwidth]{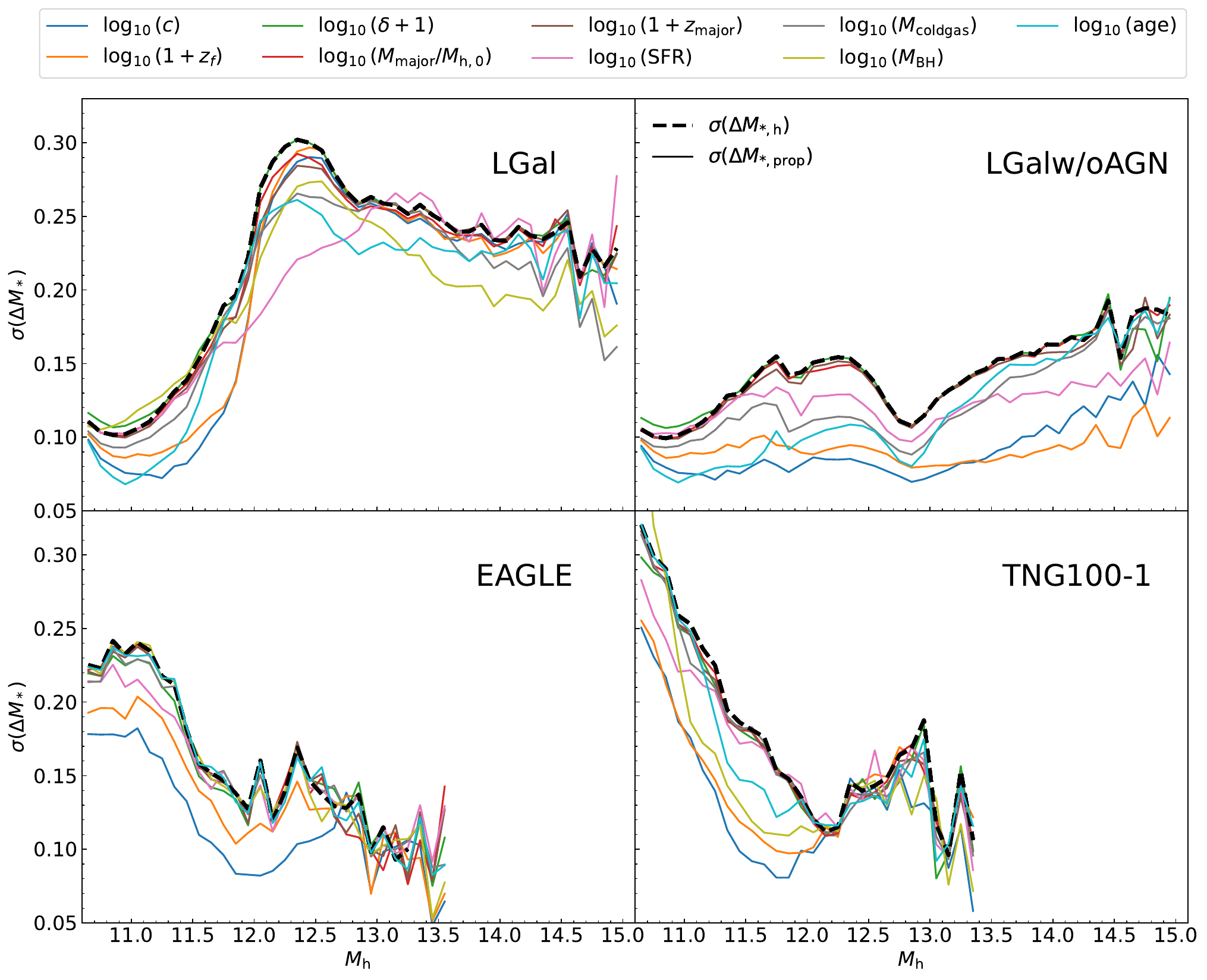}
\caption{The standard deviation of $\Delta {\rm M_*}$ from fitting formula. Black dashed lines show $\sigma(\Delta \rm M_{*, h})$ from SHMR while solid lines show $\sigma(\Delta \rm M_{*, prop})$ from the second fitting which considers secondary properties. Different panels represent different simulations as labelled in the upper-right corner. Different properties are presented by different colours labelled at the top.}
\label{fig:sigma-mvir}   
\end{figure*}

\begin{figure*}
\centering
\includegraphics[width=2\columnwidth]{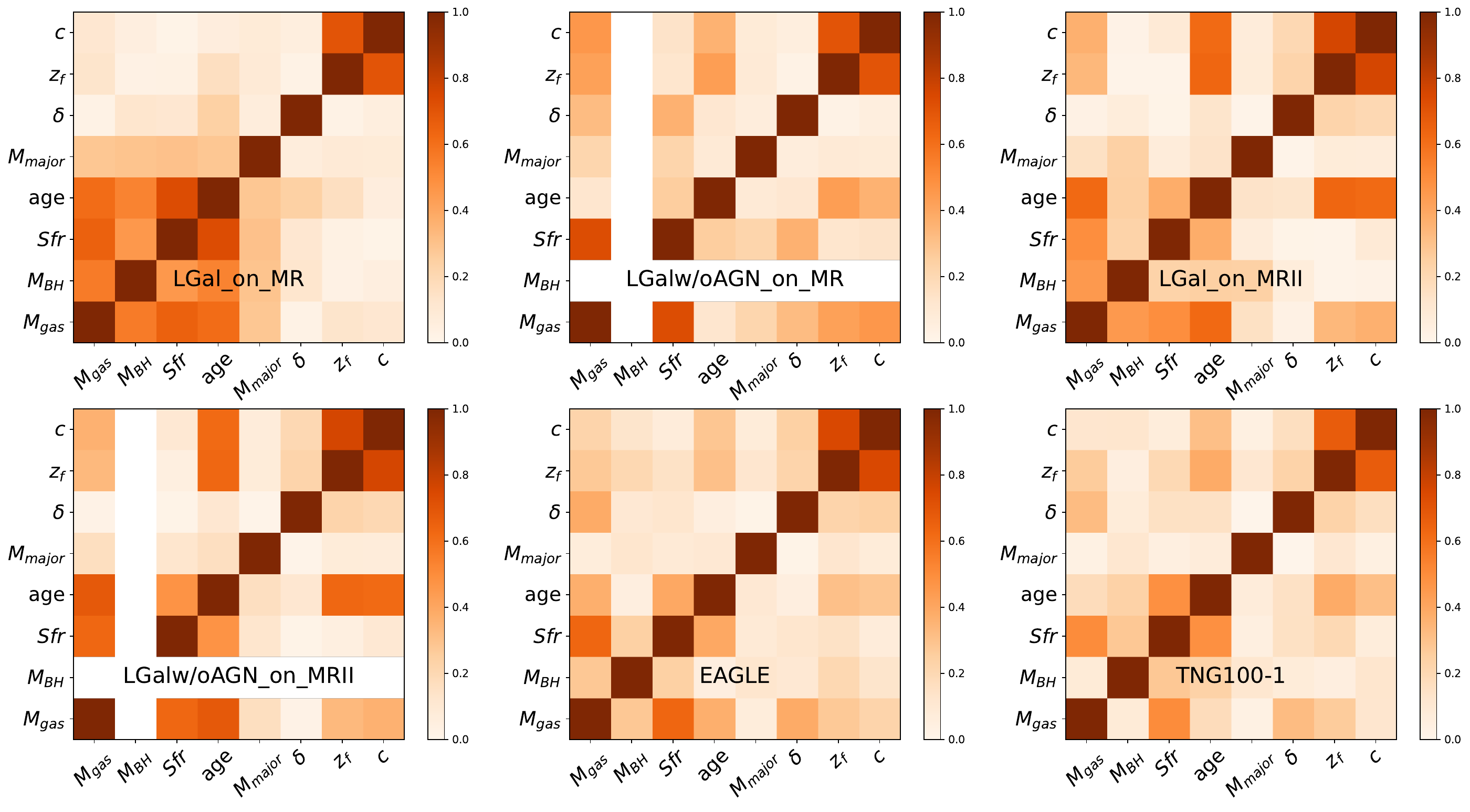}
\caption{Pearson correlation coefficients of the halo/galaxy properties used in machine learning in different simulations. The colour bar shows the absolute correlation coefficients between the two features. Larger coefficients mean stronger correlation.}
\label{fig:corr}   
\end{figure*}

\subsection{Relative dependence on halo and galaxy properties in various simulations}
\label{sec:quan}

In this section, we will first investigate the impact of individual halo and galaxy properties on the scatter around the SHMR, as well as their variations with halo masses across a range of simulations and models. Following this, machine learning techniques will be employed to evaluate the collective influence of potential halo and galaxy parameters and ascertain the importance of different properties in regulating the scatter throughout all mass categories.

\begin{figure}
\includegraphics[width=1\columnwidth]{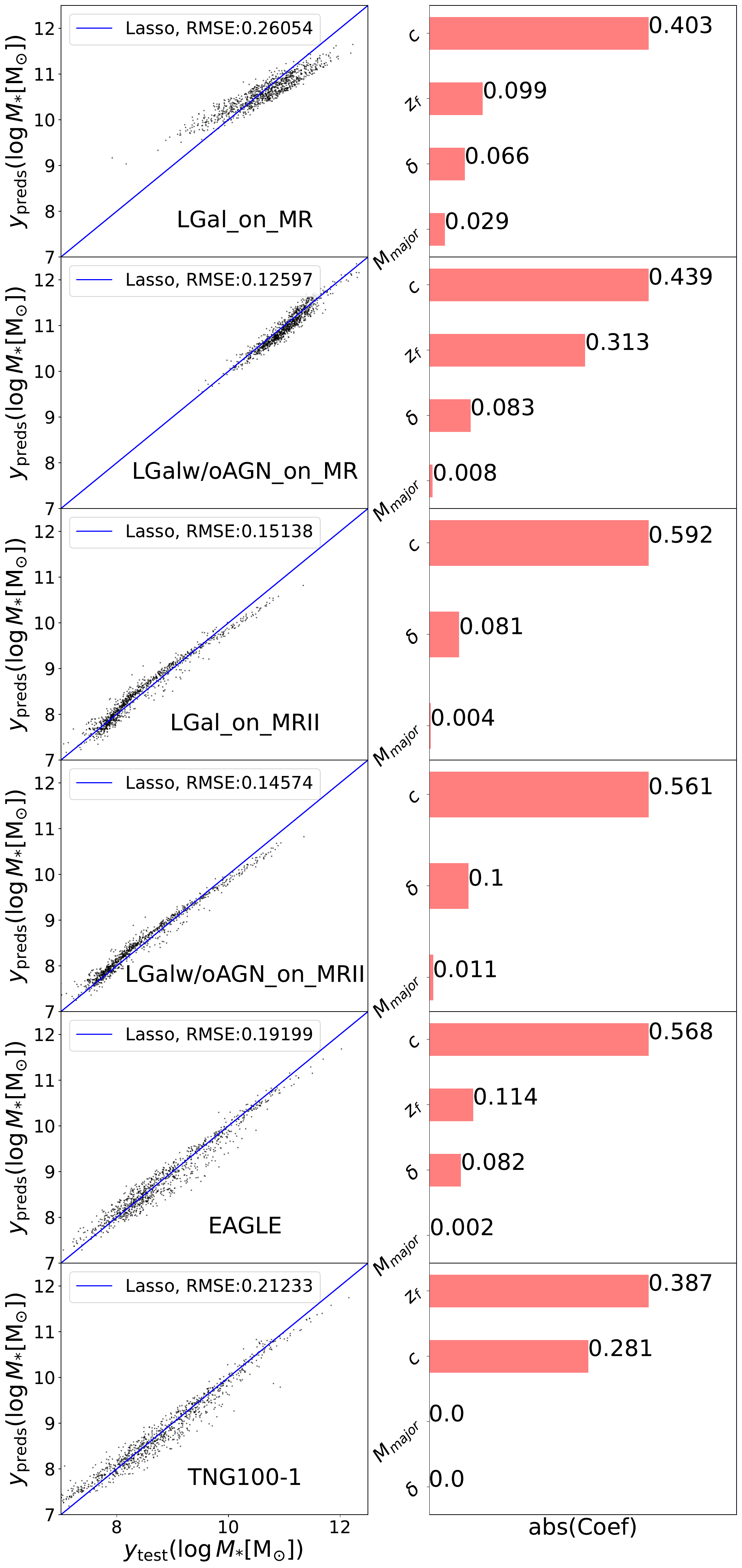}
\caption{The results of machine learning with only the halo properties as input ("Halo" model) for all the simulations. The "MR" and "MR without AGN" only contain haloes larger than $10^{12}\Msun$ while "MRII" and "MRII without AGN" only contain haloes smaller than $10^{12}\Msun$. The left column shows the predicted $M_*$ vs. true $M_*$ for the validation set. The right column shows the absolute coefficients in the result of the Lasso regression of the features we have used in the training. Since we have normalised the features into units, the coefficients represent the contributions of each feature.}
\label{fig:aah}   
\end{figure}

\begin{figure}
\includegraphics[width=1\columnwidth]{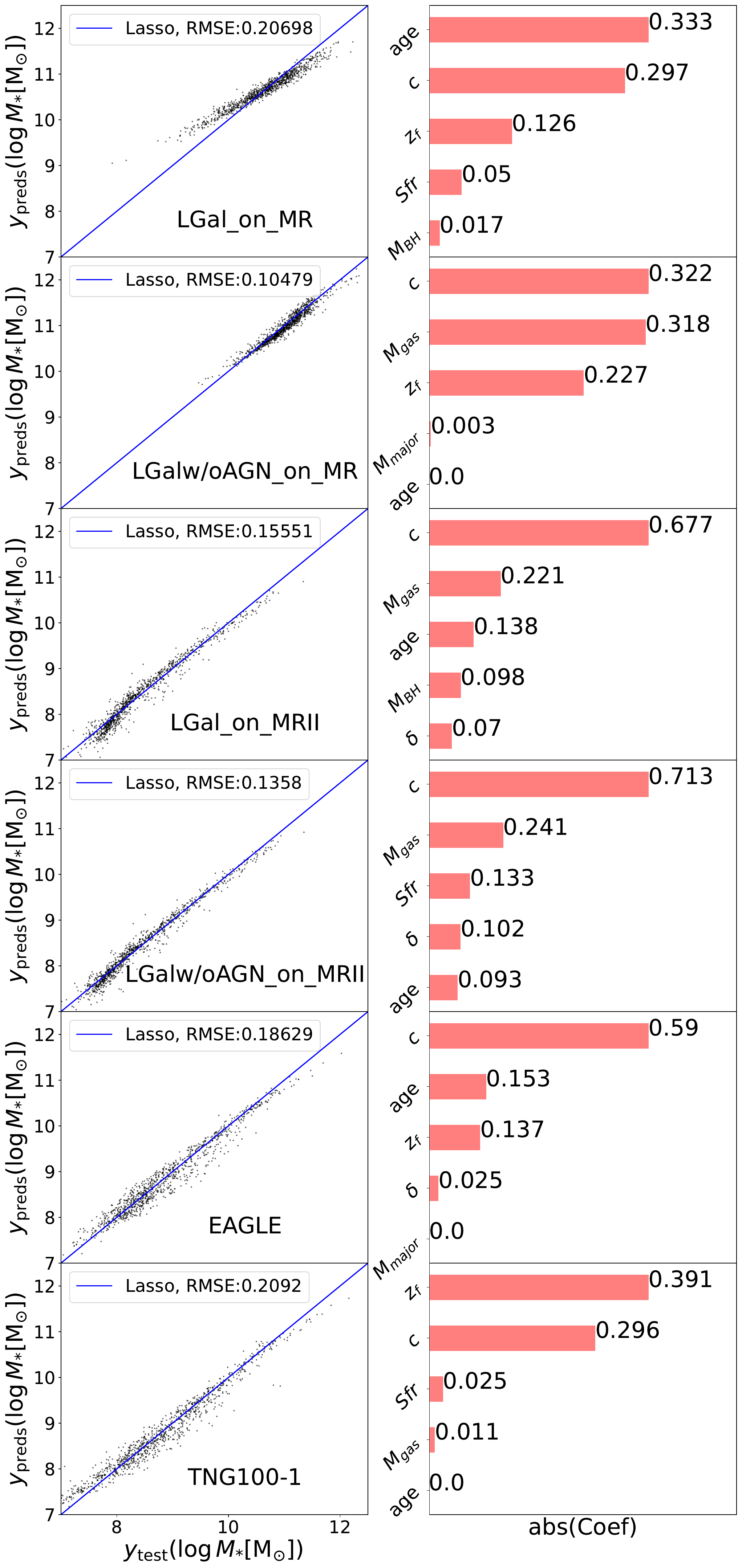}
\caption{Same as Figure~\ref{fig:aah}, but all the halo/galaxy properties are used ("All" model).}
\label{fig:aaa}   
\end{figure}

\subsubsection{Diverse impact of individual halo and galaxy properties on various halo mass levels}

Figure~\ref{fig:sigma-mvir} shows $\sigma (\Delta M_{*, \rm prop})$ as a function of halo mass for various properties. The black dashed lines represent the standard deviation of the $\Delta M_*$ (Eq.~\ref{eq:delta1}), where only halo mass is considered in determining stellar mass. Solid lines show the standard deviation of $\Delta M_{*, \rm prop}$ (Eq.~\ref{eq:delta2}), where both halo mass and secondary properties are taken into account in determining stellar mass.

The distribution of the scatter around the SHMR is influenced by both the halo properties and galaxy properties, with the extent of their contribution varying based on the halo mass and differing across different simulations. 
In \lgal{}, galaxy properties primarily influence the scatter both at low and at high masses, while halo properties such as concentration and formation time significantly affect the scatter at middle masses, $10^{11.5}\Msun<M_{\rm h} <10^{12}\Msun$. For $M_{\rm h} <10^{11}\Msun$, galaxy age exhibits a greater impact than halo concentration and formation time. In the ranges $10^{12}\Msun<M_{\rm h} <10^{12.5}\Msun$ and $10^{12.5}\Msun<M_{\rm h} <10^{13}\Msun$, SFR and age dominate the scatter, respectively. At higher masses, supermassive black hole mass (and consequently AGN feedback), becomes the driving factor. 

Since AGN feedback primarily impacts high-mass galaxies in \lgal{}, we observe that at higher masses, without AGN feedback, halo concentration and formation time are the main factors affecting the scatter around the SHMR. This highlights a co-evolution between the SFR and halo growth rate. Interestingly, at lower masses $M_{\rm h} <10^{12}\Msun$, galaxy properties like age also play a significant role in the variability in stellar mass, even in the absence of AGN feedback, showcasing the substantial influence of baryonic processes in regulating stellar mass growth in galaxies.

Significant differences are found in \eagle{} and \tng{} compared to \lgal{}. In both the \eagle{} and \tng{} simulations, it is observed that none of the properties are able to significantly diminish the scatter at high masses. At low masses, halo concentration and formation time are always seen as the primary factors contributing to scatter. A detailed comparison reveals that in \eagle{}, considering the dependence on concentration could be more effective in reducing scatter compared to formation time. Concentration remains the dominant and effective factor in reducing scatter in halos up to $10^{12.5}\Msun$. In \tng{}, the advantage of using concentration over formation time is less pronounced, and at around $10^{11}\Msun$, the dependence on concentration and formation time is similar. Additionally, concentration ceases to effectively reduce scatter in halos more massive than $10^{12}\Msun$, which is lower than in \eagle{}. In \eagle{}, no explored galaxy properties were found to effectively reduce scatter, while in \tng{}, SMBH mass, which is related to AGN feedback, could also contribute to scatter. 
By using the UM \citep{2019MNRAS.488.3143B}, \citet{2020MNRAS.493..337B} find that halo formation time is the primary contributor to the scatter in SHMR for haloes larger than $10^{13.5}\Msun$, while halo accretion rate also plays a significant role. Moreover, they note that the last major merger time contributes to the scatter, particularly for haloes with $M_{\rm h}\sim 10^{14.4}\Msun$, indicating that the halo assembly history in UM is different from \lgalaxies{}, \eagle{} and \tng{}. Additionally, \citet{2020MNRAS.491.5747M} observe that halo formation time predominantly influences the scatter in the stellar mass to peak halo mass relation, whereas environmental factors exert a weaker influence in TNG100.

In summary, it is only in the middle mass ranges, in all full physics models, that halo concentration and formation time dominate over the scatter. At low masses and higher masses, galaxy properties such as SFR, age, and SMBH mass are more closely related to the scatter in \lgal{}. In contrast, in \eagle{} and \tng{}, halo properties, especially concentration and formation time, emerge as the primary contributors to the scatter, which disappears in the most massive systems. In general, galactic properties have a greater impact on the scatter around the SHMR in \lgal{} than halo properties, as opposed to \eagle{} and \tng{}.

\subsubsection{General dependence on halo and galaxy properties}
\label{sec:ml}

In this section, we employ a machine learning approach, similar to the methodology employed by \citet{2022RAA....22h5014H}, to investigate the correlation between the scatter in the SHMR and the previously mentioned halo/galaxy attributes. For this purpose, we utilize the Lasso \citep[least absolute shrinkage and selection operator,][]{tibshirani1996regression} algorithm, which is part of Scikit-Learn \citep{2011JMLR...12.2825P}, to compute the correlation coefficient between the SHMR scatter and various features. 
Lasso is a regression analysis technique originally devised for linear models. Incorporating a penalty term proportional to the sum of the absolute values of the coefficients, the loss function is $ \frac{1}{n}\sum_{i=1}^{n}(y_i-\vec{\omega_i} \vec{x_i}-b)^2+\lambda\sum|\omega|$. It stands out for its dual functionality in variable selection and regularization. Lasso effectively shrinks less important coefficients towards zero, promoting sparsity and facilitating model interpretability. Two distinct machine-learning models have been developed, one only relying on halo properties ("Halo" model) including halo concentration $c$, halo formation time $z_{\rm f}$, overdensity $\delta$, and major merger halo mass $M_{\rm major}$, are utilized as inputs, and the other also include galaxy properties ("All" model) with star formation rate SFR, cold gas mass $M_{\rm gas}$, black hole mass $M_{\rm BH}$, and galaxy age. Note that here we exclude $z_{\rm major}$ for its influences are weak in all masses and many systems never undergo major mergers. $M_{\rm BH}$ is also excluded in simulations without AGN feedback. To evaluate the contribution of each feature to predictions, we normalize them into units. 

For each simulation, we randomly choose a sample of 20,000 central galaxies, mix them up, and split them into a training set (75\% of the sample) and a validation set (25\% of the sample). We also experiment with a different sample selection method by picking the same number of randomly chosen galaxies in each halo mass bin, with consistent results. All halo masses are considered for \eagle{} and \tng{}. For \lgal{} and \lgalwo{}, we perform machine learning in MR ($M_{\rm h} > 10^{12}\Msun$) and MRII ($10^{10.6} < M_{\rm h} < 10^{12}\Msun$) separately. These are referred to as LGal\_on\_MR, LGal\_on\_MRII, LGalw/oAGN\_on\_MR, and LGalw/oAGN\_on\_MRII, in order to maintain consistency with previous sections. 
As a consequence, a potential bias toward low masses might exist in MRII, as well as a potential bias toward high masses in MR. \red{Due to their better statistics at high masses, we place more reliance on the results from MR, despite the more pronounced resolution effect at these masses. It should be acknowledged that incorporating less massive systems in MR could somehow alter the machine learning results.}

The correlation coefficients among the eight parameters in the "All" model are presented in Figure~\ref{fig:corr}. Apart from the well-known correlation between $c$ and $z_{\rm f}$, other halo properties do not show strong correlation with each other. The coefficients among galaxy properties are higher in semi-analytical models compared to hydrodynamical simulations, indicating a degree of degeneracy in \lgalaxies{}. Particularly, the age, cold gas mass, black hole mass and SFR are strongly correlated in LGal\_on\_MR, while the connections vanish in LGal\_on\_MRII. The main reason for this are due to the different halo mass ranges adopted in MR and MRII. We have tried to explore the correlation coefficients for haloes larger than $M_{\rm h} > 10^{12}\Msun$ in LGal\_on\_MRII and find similar connections between these galaxy properties. 
To mitigate the effect of degenerate parameters, we exclude one of the features from our input if the correlation coefficient exceeds 0.75. 

Results based on the "Halo" model are presented in Figure~\ref{fig:aah}. The model exhibits satisfactory predictions overall, with better performance in MRII, \eagle{}, \tng{} and LGalw/oAGN\_on\_MR compared to those in LGal\_on\_MR with full physics, reflected by their relatively lower Root Mean Square Errors (RMSE). This aligns with expectations that at high masses galaxy growth deviates from halo growth when AGN feedback comes into play. The lower RMSE in LGal\_on\_MRII compared to hydrodynamical simulations indicates that baryonic effects impact halo properties, complexing the influence of halos on galaxy formation. The absolute coefficients in the right column show that the halo concentration and formation time are identified as the most influential features in all simulations. The environment overdensity $\delta$ has a relatively minor influence on the scatter, except for \tng{} where the coefficient drops to zero. The impact of ${M_{\rm major}}$ is negligible for all the simulations. These findings align with those presented in Figure~\ref{fig:sigma-mvir}, highlighting halo concentration and formation time as the key secondary properties that significantly reduce the scatter in the SHMR. In contrast, $\delta$ and $M_{\rm major}$ have minimal effect on reducing scatter. Furthermore, we expand our analysis to include massive haloes in MRII (all haloes with $M_{\rm h} > 10^{10.6}\Msun$) and smaller haloes in MR (all haloes with $M_{\rm h} > 10^{11}\Msun$), finding that the main outcomes of the "Halo" model remain stable, irrespective of the selected mass ranges.

Figure~\ref{fig:aaa} illustrates the results obtained using the "All" model. The overall fitting is better when including galaxy properties. The RMSE of LGal\_on\_MR decreased from 0.26 in the "Halo" model to 0.2 in the "All" model, indicating a non-negligible improvement in predicting stellar mass within high-mass dark matter haloes by incorporating galaxy properties. Other simulations show a decrease of approximately 0.02 in RMSE, representing a slight improvement in performance in the "All" model. 

In "All" model, at high masses, it is the age that dominates the scatter in LGal\_on\_MR, while concentration have a slightly smaller coefficient than age. This is consistent with those found in Figure~\ref{fig:sigma-mvir} that SFR, age, and $M_{\rm BH}$ dominate the scatter at high masses, and all these properties are strongly degenerate. For LGalw/oAGN\_on\_MR, halo concentration becomes the primary factor influencing the scatter around the SHMR, consistent with those in Figure~\ref{fig:sigma-mvir}. The galaxy property $M_{\rm gas}$ becomes a secondary important factor with a comparable effect on estimating stellar mass compared to halo concentration. 
We perform a test to train the "All" model by adding halos of relatively low mass [$10^{11}, 10^{12}$]$\Msun$ from MR (Figure~\ref{fig:aaaa}). As shown in Figure~\ref{fig:sigma-mvir}, within [$10^{11}, 10^{12}$]$\Msun$, halo formation time and concentration are key secondary properties. Considering the similar importance of halo concentration/formation time and galaxy age, the incorporation of these lower mass halos shifts the most relevant properties from galaxy characteristics to halo features, specifically the formation time. The galaxy property, $M_{\rm gas}$, has become a slightly lower but comparable effect on estimating stellar mass compared to halo formation time. In summary, galaxy properties play a dominant or subdominant role in determining the scatter around the SHMR in MR, whether excluding or including the less massive systems. Nevertheless, in both scenarios, its role is comparable to that of halo formation time and concentration. 

At low masses, halo concentration and formation time are the most efficiency properties in reducing the scatter in all the simulations, including \lgalaxies{} with and without AGN on MRII, as well as \eagle{} and \tng{}. $M_{\rm gas}$ is the second important contributor to the stellar mass but the coefficient is only 1/3 of concentration in \lgalaxies{} with and without AGN on MRII. This indicates the stellar mass in MRII is predominately determined by halo properties, and the contribution of galaxy properties is much weaker. Figure~\ref{fig:sigma-mvir} indicates that galaxy age could reduce the scatter significantly at very low masses in \lgal{} and \lgalwo{}, which is not shown in Figure~\ref{fig:aaa}. This could be attributed to the broader mass ranges used in machine learning, where halo properties are significant in the scatter at most masses. We also train the "All" model for all haloes larger than $10^{10.6}\Msun$ in MRII, and find that halo concentration remains the dominant factor in determining the scatter around the SHMR (Figure~\ref{fig:aaaa}). 

In conclusion, our machine learning analysis reveals that halo concentration/formation time dominates the scatter around the SHMR at low masses. These findings align with results obtained by \citet{2017MNRAS.465.2381M} in the EAGLE simulation and \citet{2020MNRAS.491.5747M} in TNG100. Particularly, \citet{2020MNRAS.493..337B} noted that in the UM, halo formation time contributes significantly in galaxy clusters with $M_{\rm h}\sim 10^{14.4}\Msun$, indicating that the detailed dependence relies on the underlying sub-grid physics.
Galaxy properties play a pivotal role in MR, while they contribute less in MRII. The influence of galaxy properties in hydrodynamical simulations are even weaker. In general, the role of baryonic processes varies among simulations and galaxy formation models, with baryonic processes playing a more crucial role in semi-analytical galaxy formation models in MR at high masses. 

\section{Influence of AGN feedback}
\label{sec:AGN}

One of the most significant findings of our study is the substantial decrease in the scatter of the SHMR at the high-mass end observed in \lgalaxies{} when the AGN feedback is turned off. The introduction of AGN feedback doubles the scatter at high masses, while it has no impact on the scatter at low masses. This increase in scatter is also reported by \citet{2015MNRAS.454.1038R}. At high masses, a significant number of galaxies show low stellar mass, low SFR, and low cold gas mass, indicating that AGN feedback inhibits star formation activity in these galaxies, resulting in the appearance of a population of low-mass galaxies that contribute significantly to the scatter. This is in line with the conclusions of \citet{2021NatAs...5.1069C} in the SIMBA simulation, where jet-mode and X-ray AGN feedback suppresses galaxies, leading to the formation of red, lower-mass galaxies.

Incorporating AGN feedback additionally decreases the influence of factors such as halo concentration and formation time on the scatter around the SHMR. At lower masses ($<10^{12} \Msun$) where AGN feedback is ineffective in all simulations and models, star formation activity aligns with halo growth. Consequently, halo concentration and formation time serve as secondary factors in determining stellar mass (major contributors to the scatter around the SHMR). In contrast, at higher masses, AGN feedback impacts the entire star formation process, decreasing the influence of halo concentration and formation time in \eagle{} and \tng{}. The AGN feedback effect is more prominent in \lgal{}, becoming the main driver and resulting in an increase in the scatter around the SHMR.

In summary, AGN significantly affects not only the scatter amplitude but also the impact of halo properties on determining the scatter around the SHMR, especially at high masses. While models to link galaxies and halos using halo mass and concentration are reliable at low masses, more caution should be exercised in establishing such a link in massive systems.

\section{Conclusion}
\label{sec:conclusion}

 Despite a well-established correlation between the stellar mass and halo mass, substantial variations in stellar mass persist at a given halo mass. In this work, we analyze data from three simulations—\lgalaxies{}, \eagle{}, and \tng{}—we explore a range of halo properties, encompassing halo concentration, formation time, environment, and mass assembly history. Additionally, we consider galaxy properties such as star formation rate, age, cold gas mass and supermassive black hole mass. To discern the influence of AGN feedback, we compare results from simulations with and without this process. Overall, our findings exhibit consistency across simulations, with minor divergences attributed to distinct sub-grid physical recipes. Our main conclusions can be summarised as follows:
\\[-.55cm]
\begin{enumerate}
 	\item For low-mass haloes ($M_{\rm vir} < 10^{11.5}\Msun$), the scatter in SHMR is almost constant ($\sigma\approx0.15$) in \lgalaxies{}, while the scatters are much larger in \eagle{} and \tng{} simulations. For high-mass haloes ($M_{\rm vir} > 10^{12}\Msun$), the scatter is 2 times larger in \lgal{} (with AGN feedback on) compared to \lgalwo{} (with AGN feedback off), \eagle{}, and \tng{}.
        \\[-.3cm]
        \item The scatter in the SHMR strongly depends on the concentration and the formation time of the host haloes, especially at lower masses. For a given halo mass, galaxies in early-formed high-concentration haloes exhibit an excess of stellar mass compared to those in late-formed low-concentration haloes.
        \\[-.3cm]
        \item Baryonic processes play a more critical role in determining the scatter around the SHMR in \lgal{} compared to \eagle{} and \tng{}. In \lgal{}, SFR, age, and BH mass dominate the scatter around the SHMR in most halo mass ranges except for $10^{11.5}\Msun<M_{\rm h}<10^{12}\Msun$, where halo concentration and formation time take precedence. Conversely, in \eagle{} and \tng{}, halo concentration and formation time govern the scatter at lower masses, with galaxy properties never being the primary driver of the scatter.
        \\[-.3cm]
        \item AGN feedback plays a crucial role in the scatter around the SHMR at high masses. It dominates the scatter in \lgal{} and diminishes the dependencies of scatter on halo concentration and formation time in \eagle{} and \tng{}.
        \\[-.3cm]
        \item Environmental overdensities and mergers do not significantly affect the scatter in the SHMR. Their impact is noticeable only in low-mass halos. Galaxies in high-density areas have slightly higher stellar masses than those in low-density areas; galaxies in halos that experienced an early major merger have slightly larger stellar masses compared to galaxies in halos with a late major merger.
 	
\end{enumerate}

In summary, our analysis reveals that the source of the scatter in SHMR exhibits a transitional mass at around $10^{12}\Msun$ and varies among simulations and models. At lower masses, more massive galaxies are characterized by higher halo concentration, earlier formation time, denser environment, and earlier last major merger time given a halo mass. In \lgal{}, SFR and age are more important than in \eagle{} and \tng{}. At higher masses, AGN feedback plays a prominent role in increasing the scatter, and in diminishing the influence of halo concentration and formation time on the growth of stellar mass. In general, in semi-analytical models, baryonic processes play a more important role compared to \tng{} and \eagle{}. 

\section*{Acknowledgements}

This work is supported by the National SKA Program of China (Nos. 2022SKA0110201 and 2022SKA0110200), and CAS Project for Young Scientists in Basic Research grant No. YSBR-062, the National Natural Science Foundation of China (NSFC) (grant Nos. 12033008, 12273053, 11988101, 11622325), the K.C.Wong Education Foundation and the science research grants from China Manned Space Project with No.CMS-CSST-2021-A03 and No.CMS-CSST-2021-B03.

\section*{DATA AVAILABILITY}
The data produced in this paper are available upon reasonable request to the corresponding author.




\bibliographystyle{mnras}
\bibliography{ref} 






\appendix

\section{SHMR in different simulations}
\label{app:shmr}

\begin{figure*}
\centering
\includegraphics[width=1.6\columnwidth]{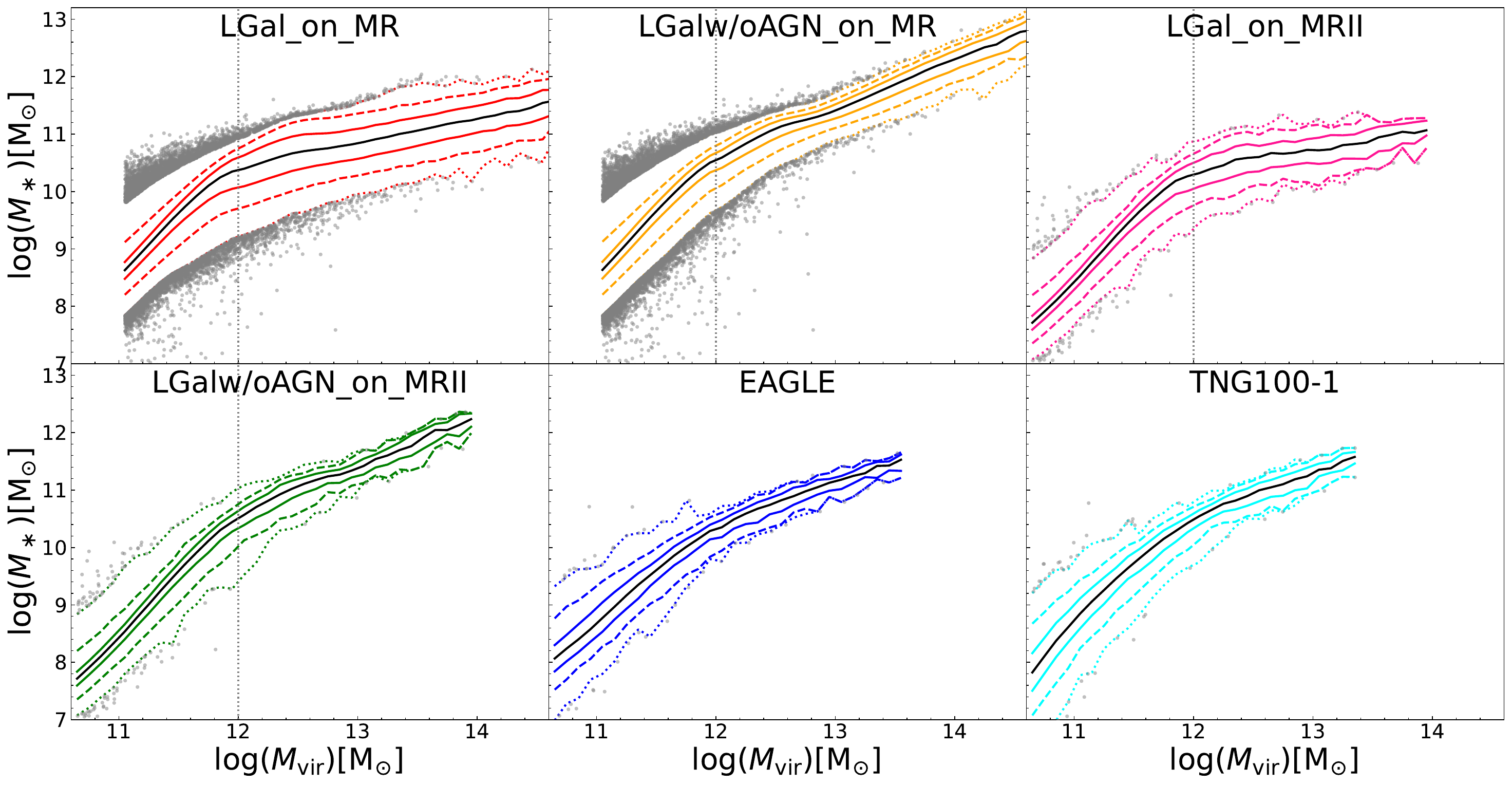}
\caption{The SHMR in different simulations. We show the results from MR and MRII separately. Different panels represent different simulations, while the black solid lines represent the median value, the solid/dashed/dotted coloured lines in each panel represent the 68\%/96\%/99.8\% regions, respectively. The vertical dashed lines show the point where $M_{\rm h} = 10^{12}\Msun$. Grey points denote galaxies that fall beyond the 3$\sigma$ boundaries.
}
\label{fig:smhm_sca}
\end{figure*} 

To illustrate the SHMR and the scatter in different simulations better, we further show the SHMR in Figure~\ref{fig:smhm_sca}. Different panels represent different simulations, while the black solid lines represent the median value, the solid/dashed/dotted coloured lines in each panel represent the 68\%/96\%/99.8\% regions, respectively. Grey points denote galaxies that fall beyond the 3$\sigma$ boundaries. Here we show the results from MR and MRII separately, and select galaxies from MR with $M_{\rm h} > 10^{11}\Msun$ and from MRII with $M_{\rm h} > 10^{10.6}\Msun$. The vertical dashed lines show the point where $M_{\rm h} = 10^{12}\Msun$. We find that at mass range $10^{11}<M_{\rm h}<10^{12}\Msun$, both LGal\_on\_MR and LGalw/oAGN\_on\_MR show similar 2$\sigma$ scatter as LGal\_on\_MRII and LGalw/oAGN\_on\_MRII, indicating that the \lgalaxies{} has a good convergence at this mass range. The outlier is much larger in MR than in MRII due to the larger sample sizes in MR. However, LGal\_on\_MR shows a steep increase around $10^{12}\Msun$, which is absent in the other three simulations with \lgalaxies{}. This could be attributed to different SMBH growth histories which are closely related to mergers and thus sensitive to the resolution of the simulation. The good overall agreement between LGalw/oAGN\_on\_MR and LGalw/oAGN\_on\_MRII also indicates that the increasing scatter in LGal\_on\_MR is related to SMBH and AGN feedback. Both \eagle{} and \tng{} show a decreasing scatter with increasing halo mass. At low masses, they predict higher scatter than \lgal{}. In general, the overall behaviour of the scatter is similar to Figure~\ref{fig:smhm}.

\section{The dependence on halo formation time and last major merger time}

\begin{figure*}
\centering
\subfloat{\includegraphics[width=1.6\columnwidth]{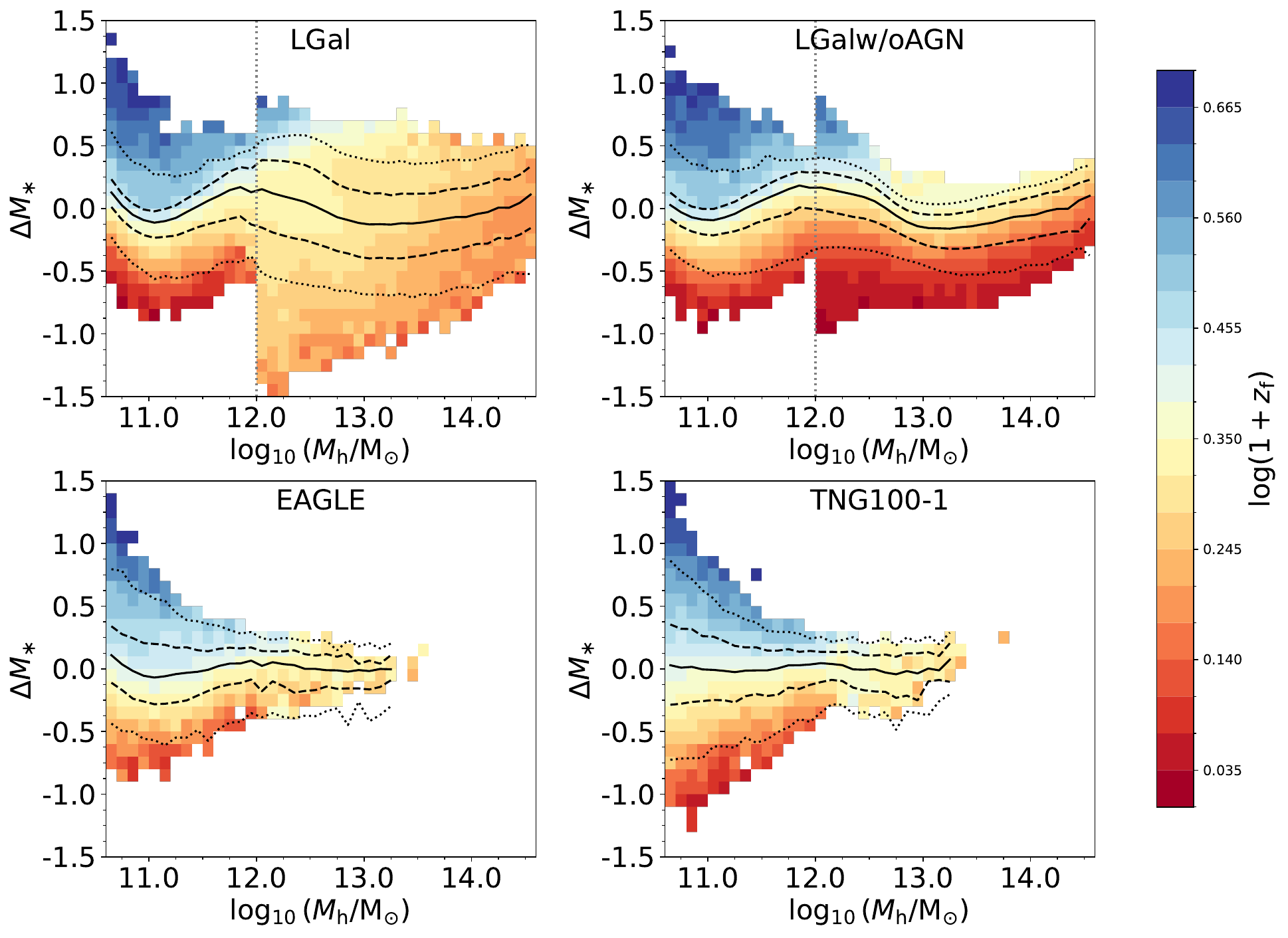}}\\
\subfloat{\includegraphics[width=1.6\columnwidth]{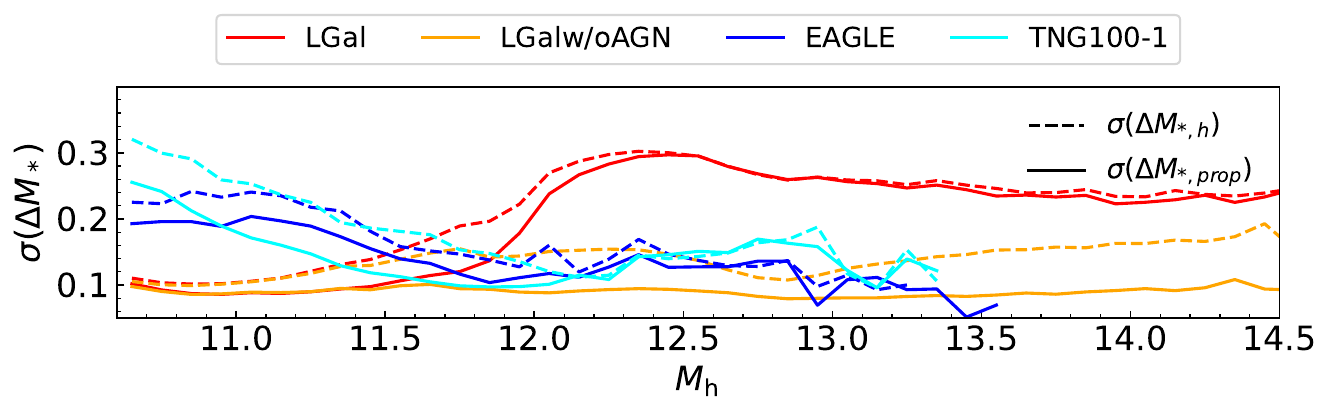}}
\caption{Same as Figure~\ref{fig:c} but each bin is coloured by the formation time, $z_{\rm f}$, of the haloes. Solid lines in the bottom panel show $\sigma (\Delta \rm M_{*, prop})$ from the second fitting which considers $z_{\rm f}$.}
\label{fig:tf}
\end{figure*}

\begin{figure*}
\centering
\subfloat{\includegraphics[width=1.6\columnwidth]{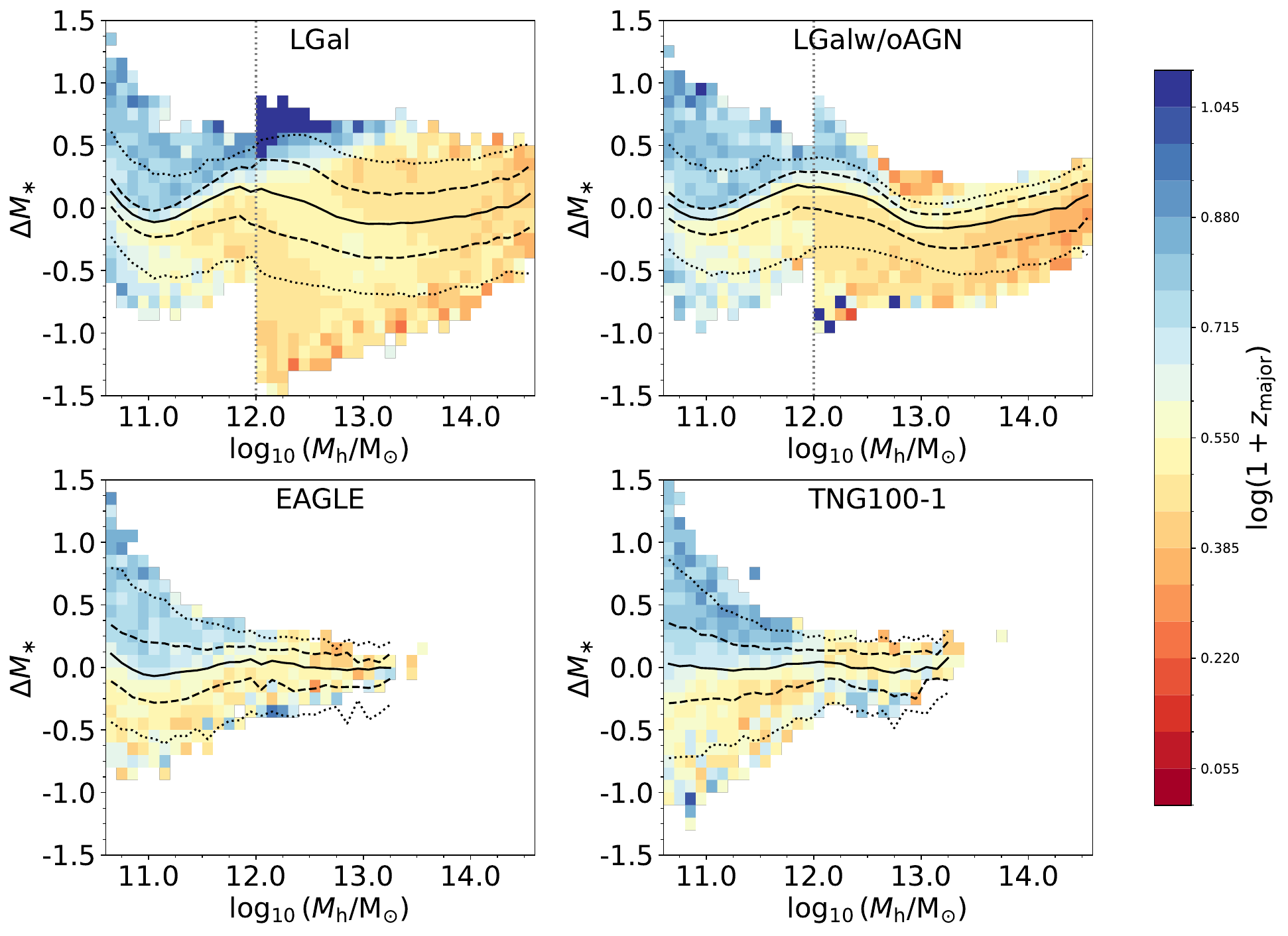}}\\
\subfloat{\includegraphics[width=1.6\columnwidth]{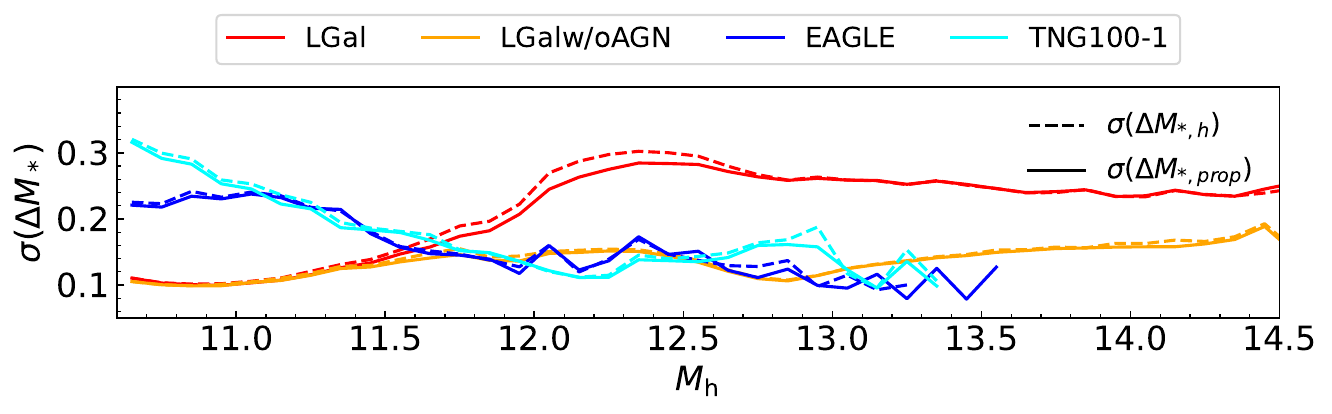}}
\caption{Same as Figure~\ref{fig:c} but is coloured by the mean redshift of the last major merger, $z_{\rm major}$. Solid lines in the bottom panel show $\sigma (\Delta \rm M_{*, prop})$ from the second fitting which considers $z_{\rm major}$.}
\label{fig:lmt}
\end{figure*}

\subsection{Halo formation time}
\label{sec:tf}
The halo formation time is tightly related to its concentration, as early-formed halos are more concentrated \citep{2002ApJ...568...52W, 2009ApJ...707..354Z, 2011MNRAS.415L..69J, 2014MNRAS.441..378L,2015MNRAS.450.1521C}. Our works also find that the correlation between the scatter in SHMR and halo formation time resembles that of concentration. The $\Delta M_{\ast}$-$M_{\rm h}$ relation, colour-coded by halo formation time, is shown in Figure~\ref{fig:tf}, similar to Figure~\ref{fig:c}. Galaxies in earlier-formed halos tend to be more massive, with this trend being particularly pronounced at lower masses across all these simulations. At high masses, the trend is relatively weak except for \lgalwo{}, indicating that AGN feedback may decouple the growth of stellar components from halo accretion. All these findings mirror the dependence on concentration, revealing the strong correlation between halo formation time and concentration. Further discussion and comparison with prior studies are provided in Section~\ref{sec:c}.

The bottom panel in Figure~\ref{fig:tf} quantifies the dependence on halo formation time as a function of host halo masses. At low masses, the scatter is reduced by approximately one-third after considering formation time. However, at high masses, the role of $z_{\rm f}$ weakens, with AGN influence becoming predominant. Particularly, for \lgalwo{}, the scatter can be reduced by 30\% across all mass ranges. The overall dependence on formation time mirrors that on concentration (illustrated in the lower panel of Figure~\ref{fig:c}), affirming the robust correlation between halo concentration and formation time. Furthermore, it is noteworthy that the magnitude of the decrease in formation time is slightly smaller compared to concentration, especially for \lgal{} and \eagle{}, suggesting that concentration may have a greater influence on the scatter than formation time.

\subsection{Last major merger time}
\label{app:lmt}
We also consider the last major merger redshift, $z_{\rm major}$, to further investigate the impact of major mergers. For a central galaxy, $z_{\rm major}$ represents the redshift when it underwent its last major merger. A larger $z_{\rm major}$ indicates that the galaxy experienced its last major merger earlier and has had a relatively smooth growth history since then. In Figure~\ref{fig:lmt}, we present the SHMR coloured according to $z_{\rm major}$. If a halo does not experience a major merger in the given simulation, we set $z_{\rm major}$=127. The impact of $z_{\rm major}$ is similar across all simulations at low masses. Massive galaxies with similar halo mass have a higher $z_{\rm major}$, implying that galaxies with a more gradual growth history accumulate more stellar mass. This is also related to the positive correlation between $z_{\rm major}$ and halo formation time. The influence of $z_{\rm major}$ weakens significantly at high masses. Similar to the halo concentration and formation time, at these masses, AGN feedback can suppress star formation, while the host haloes keep growing, uncoupling the stellar mass growth history from the halo mass growth history.

We show the quantitative impact of $z_{\rm major}$ in the bottom panel of Figure~\ref{fig:lmt}. At all masses, the contribution of $z_{\rm major}$ is almost negligible. This influence is slightly larger at $10^{12}<M_{\rm h}<10^{12.5}\Msun$ and can reach up to 10\% in \lgal{}, which diminishes towards both lower and higher halo masses.

\section{Dependence on box size and resolution in hydrodynamical simulations}
\label{app:conv}

\begin{figure*}
\centering
\includegraphics[width=1.6\columnwidth]{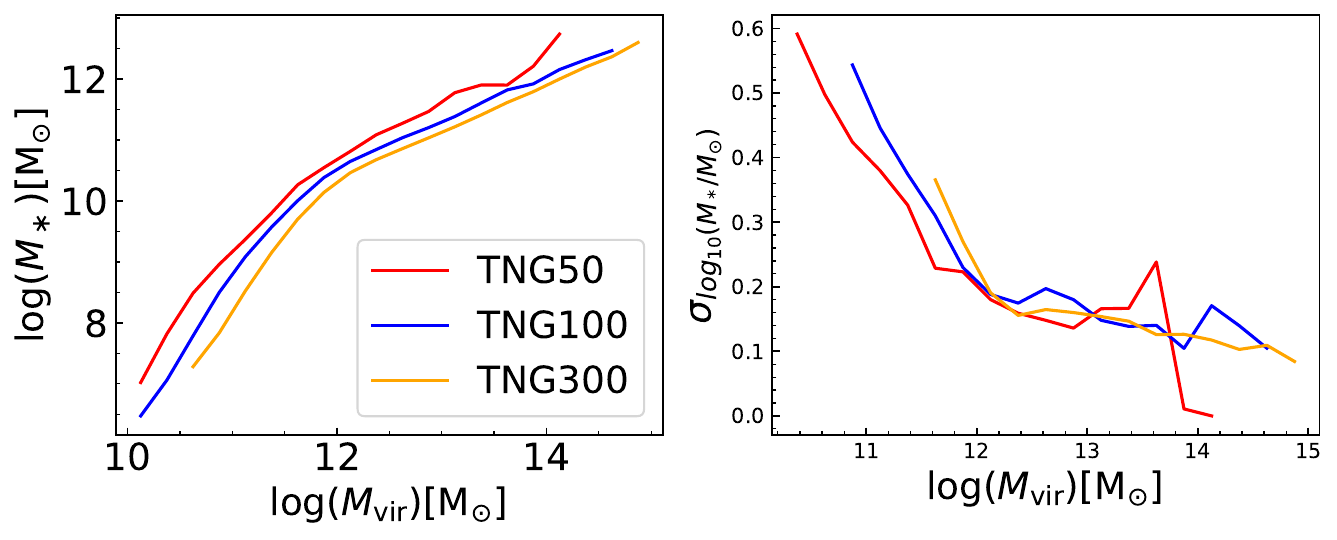}
\caption{Left panel: the median SHMR in TNG50 (red), TNG100 (blue) and TNG300 (yellow). Right panel: the 1$\sigma$ scatter of the SHMR in TNG simulations.}
\label{fig:tngconv}
\end{figure*}

It is worth noting that \textit{IllustrisTNG} might have some convergence problems in simultaneously reproducing the stellar mass function with the same sub-grid physics and parameters in simulations with different box sizes and particle masses, like TNG100 and TNG300 \citep{2018MNRAS.475..648P}. In this section, we test whether the scatter in SHMR as a function of halo mass is influenced by such a phenomenon. We examine the scatter in TNG50, TNG100, and TNG300 simulations and present the results in Figure~\ref{fig:tngconv}. The left panel shows the SHMR in these simulations. We observe an increasing stellar mass with decreasing box size at fixed halo mass, indicating that \textit{IllustrisTNG} tends to predict more massive galaxies with smaller box sizes and higher resolutions. The right panel shows the scatter in SHMR as a function of halo mass in these simulations. We find a general decreasing trend towards the higher halo mass. However, simulation with smaller box, higher resolution predicts lower scatter at fixed halo mass with $M_{\rm h} < 10^{12}$. The scatter appears to be independent of the size of the simulation box at high masses. For the \eagle{} simulation, \citet{2017MNRAS.465.2381M} compared the scatter in EAGLE100 and EAGLE50 and found similar results (see their Appendix B for details). They also report that higher resolution tends to have a smaller scatter in SHMR.

\section{Machine Learning on the whole mass range of MR and MRII}

\begin{figure*}
\centering
\includegraphics[width=2\columnwidth]{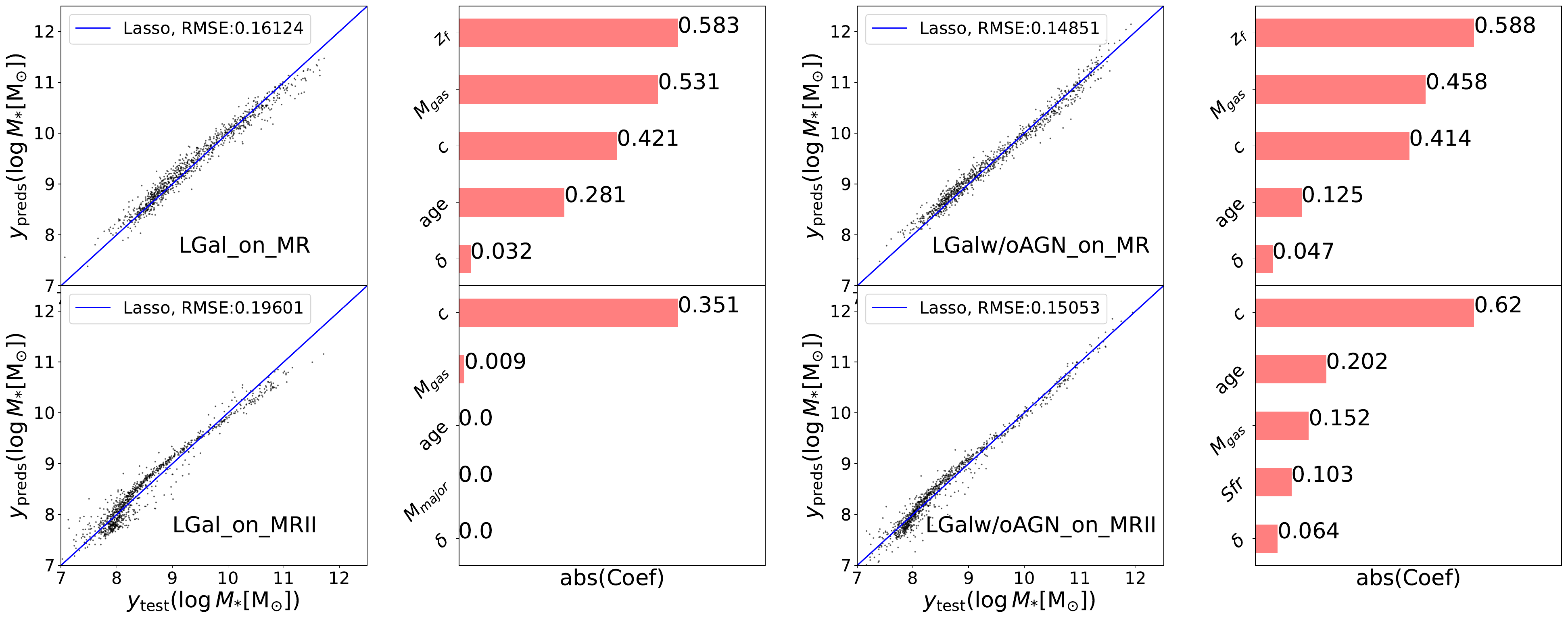}
\caption{Same as Figure~\ref{fig:aaa}, but all the available halo mass ranges ($M_{\rm h} > 10^{11}\Msun$ for MR and $M_{\rm h} > 10^{10.6}\Msun$ for MRII) are used in "All" model.}
\label{fig:aaaa}   
\end{figure*}

We further expand our machine learning method to include massive haloes in MRII (all haloes with $M_{\rm h} > 10^{10.6}\Msun$) and smaller haloes in MR (all haloes with $M_{\rm h} > 10^{11}\Msun$) for both "Halo" model and "All" model. The main findings in the "Halo" model remain stable, irrespective of the selected mass ranges. The main results from "All" model are also consistent with the one presented in Section~\ref{sec:ml}, with slightly differences as shown in Figure~\ref{fig:aaaa}. We find that galaxy properties play a vital important role compared to halo properties in MR, while the scatter are determined by halo concentration along in MRII. 



\bsp	
\label{lastpage}
\end{document}